\documentclass[aps,prd,twocolumn,superscriptaddress]{revtex4-2}
\usepackage{amsmath}
\usepackage{graphicx}% Include figure files
\usepackage{dcolumn}% Align table columns on decimal point
\usepackage{bm}% bold math
\usepackage{float} % allows [H] placement to force exact position
\usepackage{hyperref} %gives hyperlink to refs
\usepackage{xcolor} % more color options
\usepackage[caption=false]{subfig}

\usepackage{placeins}
 \usepackage{pifont,xcolor}
 \newcommand{\cmark}{{\color{green!60!black}\ding{51}}}
 \newcommand{\xmark}{{\color{red!70!black}\ding{55}}}

\usepackage{xcolor}

\definecolor{darkgreen}{rgb}{0.0,0.45,0.0}

\begin{document}

\preprint{APS/123-QED}

\title{Nonlinear Stability of Kerr--Sen Black Holes in Merging Binaries}% Force line breaks with \\

\author{Andrew Carroll}
\email{andrew.ajc.carroll@gmail.com}
\affiliation{Department of Physics and Astronomy, Brigham Young University, Provo, Utah 84602, USA}

\author{Eric W. Hirschmann}
\email{ehirsch@byu.edu}
\affiliation{Department of Physics and Astronomy, Brigham Young University, Provo, Utah 84602, USA}

\author{Hyun Lim}
\email{hyunlim@lanl.gov}
\affiliation{Applied Computer Science (CAI-1) and Center for Theoretical Astrophysics, Los Alamos National Laboratory, Los Alamos, New Mexico 87545, USA}

\author{David Neilsen}
\email{david.neilsen@byu.edu}
\affiliation{Department of Physics and Astronomy, Brigham Young University, Provo, Utah 84602, USA}

\author{David F. Van Komen}
\email{david.vankomen@gmail.com}
\affiliation{Kahlert School of Computing, University of Utah, Salt Lake City, Utah 84112, USA}

\author{Sebastian Vander Ploeg Fallon}
\email{sjvpf1@gmail.com}
\affiliation{Department of Physics, Cornell University, Ithaca, New York 14853, USA}

\date{\today}

\begin{abstract}
We investigate the stability of Kerr–Sen black holes, which arise in Einstein-Maxwell-dilaton-axion theory.   
Within a numerical relativity framework, we perform head-on binary black hole simulations with approximate initial data across a portion of the  parameter space. %We calculate the radiation extracted in gravitational, electromagnetic and scalar channels.
We find that for nontrivial electric charge, a dilaton field persists through merger and that in the presence of spin, the remnant will also retain an axion field. The persistence of these fields for long times after merger strongly suggests the stability of these black holes within this alternative gravity
theory.
We further test whether initially unscalarized black holes will acquire hair in the presence of a scalar background. We find that black holes immersed in such a background retain scalar hair. Furthermore, we find that even initially unscalarized Kerr–Newman black holes will scalarize and remain scalarized throughout the evolution.

\end{abstract}
\maketitle
 
%\tableofcontents

\section{Introduction}
\label{sec:intro}

While general relativity (GR) has passed all current experimental tests, there exist a large number of potentially viable extensions to it. Ongoing improvements in detector sensitivity, combined with numerical relativity will continue to enable increasingly precise tests of GR and proposed alternatives~\cite{Abbott2025GWTC3GR,Yunes2025GWTests,Abac2025GW250114,
Giri2025AdSBlackShells,Clough2024,Bambi2025Mimickers,
Yunes2013_LRR_GRTests,Yagi2016_BHTestsGR,Meidam2014,Berti2015,Yang2017}.
Indeed, advances in gravitational wave (GW) astronomy have led to a dramatic increase in the total number of detections of
signals from compact binary mergers, thereby providing access to strongly
gravitating, nonlinear domains. Since the first detection of gravitational waves, 
the Advanced LIGO and Virgo detectors have completed multiple
observing runs, producing a growing catalog of compact binary mergers
\cite{Abbott2019,GWTC2,GWTC3,GWTC4}. All observed mergers to date are consistent
with the predictions of GR within current uncertainties~\cite{Abbott2016,Abbott2016b,Abbott2019}. Nevertheless, compact-binary coalescences continue to provide one of the most promising arenas for testing gravity beyond GR.

The increasing precision of gravitational wave observations has motivated effort to model compact object dynamics in theories beyond general relativity~\cite{Cardoso2012NRHEP,Corman2024NonlinearModificationsGR,Afshordi2024BlackHolesInsideOut,Bambi2025BlackHoleMimickers,Bernard2025GenericEFTGWTests}. Numerical relativity is
essential for this program because the late inspiral, merger, and ringdown
probe regimes in which the dynamics are intrinsically nonlinear and
analytic approximations are not uniformly reliable. Extending numerical
relativity to alternative theories of gravity, however, requires careful
control of the associated initial-value problem. Reliable nonlinear
simulations require evolution systems with suitable hyperbolic structure,
controlled constraint propagation, and stable gauge dynamics. These
considerations have become central in beyond-GR numerical relativity,
where the construction of robust evolution formulations is often a
prerequisite for extracting physical predictions.

The challenge is particularly relevant for theories motivated by effective field theory or higher curvature corrections to GR. Such theories introduce higher derivatives, additional characteristic modes, or principal parts whose hyperbolicity depends on the background solution and coupling strength. 
In some cases, higher derivatives may also be associated with Ostrogradsky-type instabilities, unless the theory is interpreted within a controlled effective description or reformulated appropriately.
A significant number of previous studies have therefore been devoted to approaches that permit a more complete consideration of nonlinear aspects while maintaining control over the fundamental evolutionary problem. Examples of these include order reduction schemes in dynamical Chern-Simons gravity~\cite{Alexander2009, Okounkova2019} and Einstein-dilaton-Gauss-Bonnet theory~\cite{Okounkova:2020rqw,Okounkova2020}, modified gauge choices in Einstein-scalar-Gauss-Bonnet~\cite{Ripley2020a,Ripley2020b}, the so-called ``fixing the equations" approach~\cite{Cayuso2020}, and auxiliary field formulations for solving higher curvature systems including quadratic gravity~\cite{Held2021,Held2023,Held2025,East2023,Taillte2025} (see also~\cite{Corman2024} for a more in-depth discussion and comparison of these and other approaches). These developments have expanded the range of beyond-GR theories accessible to numerical relativity while emphasizing that dynamical consistency and observational phenomenology must be addressed together.

From this perspective, theories whose field equations can be written as second order evolution systems provide particularly useful laboratories for beyond-GR numerical relativity. They allow one to study additional propagating degrees of freedom in fully nonlinear black hole (BH) spacetimes without relying entirely on perturbative or order reduced treatments of the gravitational field equations. 
Einstein--Maxwell--Dilaton--Axion (EMDA)~\cite{Sen1992,Sen1995}, arising as a low-energy limit of heterotic string theory, is one such theory well suited to this purpose. Unlike many higher curvature modifications to GR, EMDA does not introduce higher derivatives into the metric field equations. Instead, it extends the Einstein-Maxwell system by coupling the electromagnetic field to two additional propagating matter degrees of freedom: a dilaton and an axion. The resulting equations retain a structure closely related to the Einstein-Maxwell system coupled to nonlinear matter fields. Consequently, EMDA can be formulated as a second order hyperbolic evolution problem using standard numerical relativity methods, together with appropriate evolution equations for the electromagnetic, dilaton, and axion fields.  In the limit of vanishing axion coupling, EMDA reduces to Einstein--Maxwell--Dilaton (EMD) theory.

Having a well posed initial value problem for EMDA is a crucial ingredient for sensible simulations, but it is also advantageous as it has analytically known black hole solutions at specific points in the theory space.  Certain of these are the stationary Kerr--Sen black holes.  They generalize the Kerr--Newman family of general relativity in that they carry 
nontrivial dilaton and axion fields in addition to mass, angular momentum,
and electromagnetic charge. As is well known, in standard GR coupled to Maxwell fields,
stationary black holes are characterized by their mass, angular momentum,
and electric charge \cite{Israel1967,Carter1971}. In contrast, extensions
of GR such as EMDA can admit black holes with additional matter fields.
Kerr--Sen solutions therefore
provide a natural setting in which to explore whether these additional degrees of
freedom behave as stable black hole hair in the nonlinear strong field
regime. Related questions have been studied in scalar-Gauss--Bonnet gravity,
where scalarization and dynamical descalarization can occur during
compact binary evolutions
\cite{Elley2022,Silva2021DynamicalDescalarization}.
The nonlinear behavior of scalar and axionic hair in EMDA is connected to
several lines of previous work. Studies of EMD~\cite{Julie2017,Hirschmann2018}
provide an important foundation for extending nonlinear simulations to the
full EMDA system. Analytical investigations have examined aspects of the
perturbative stability of EMDA black holes in certain parameter regimes
\cite{PopeRohrerWhiting2025}. In addition, astrophysical studies have
explored possible observational signatures of Kerr--Sen BHs. For example,
analyses of Event Horizon Telescope (EHT) shadow data suggest that Sgr A*
may favor the Kerr--Sen solution, although similar data from M87* favor
Kerr while remaining consistent with Kerr--Sen within current observational
uncertainties~\cite{Sahoo2024PRD,Sahoo2025}.

Fully nonlinear simulations of binary BHs in EMDA remain limited. In particular, it is not yet well understood whether dilaton and axion hair persist through merger, disperse via radiation, or settle into a stable remnant configuration. Addressing these questions is important for assessing the nonlinear stability of Kerr--Sen BHs and for determining whether EMDA can provide distinct strong field signatures for future GW observations.  Although significant progress has been made in waveform modeling for selected beyond-GR theories, including scalar-Gauss-Bonnet and Einstein-dilaton-Gauss-Bonnet gravity, waveform coverage in alternative theories remains far less developed than in GR. Consequently, many current tests still rely on phenomenological parameterizations, post-Newtonian approximations, or targeted numerical relativity simulations~\cite{Gasparotto2026MemoryHairyBBH,Agathos2014,Elley2022,Doneva2023sGB}.

In this study, we extend these investigations by simulating head-on black hole mergers governed by the full EMDA equations. We analyze the resulting waveforms and examine the stability of the remnants, thereby gaining insight into the persistence of scalar hair within the theory. Specifically, we evolve several head-on collisions of BHs in EMDA to determine the conditions under which the dilaton and axion fields survive merger.

The remainder of this paper is organized as follows. In Sec.~\ref{sec:emda}, we review the EMDA action, equations of motion, and the formulation used for numerical evolution. In Sec.~\ref{sec:methods}, we describe our numerical methods, including the modified \texttt{Dendro-GR} infrastructure, the construction of approximate binary BH initial data, and the extraction of gravitational, electromagnetic, dilaton, and axion radiation. In Sec.~\ref{sec:results}, we present simulations of head-on Kerr--Sen BH collisions, analyze the emitted radiation, and examine the persistence of scalar hair in the remnant. We also study initially unscalarized configurations to determine whether scalar hair can be generated dynamically. We conclude in Sec.~\ref{sec:conclusion} with a summary of our results and directions for future work. Several important details such as the equations of motion, the analytic form of the Kerr--Sen solution, initial data and wave extraction techniques and issues associated with code performance are relegated to a number of appendices in order not to disrupt the continuity of the discussion.  Throughout this paper we use a mostly positive signature, Latin letters as spacetime indices, and work in geometrized units, $G=c=1$. %When discussing binary configurations, we use lowercase $m$ and $q$ to denote the mass and charge of an individual black hole, while uppercase $M$and $Q$ denote the mass and charge associated with the full spacetime of the binary system.

\section{Einstein--Maxwell--Dilaton--Axion}
\label{sec:emda}
The general theory we consider in this work is an extension of Einstein-Maxwell on inclusion of two coupled scalar fields, the dilaton and the axion.  The action for the EMDA system is
\begin{equation}
\begin{aligned}
S=\int \mathrm{d}^4 x\,\sqrt{-g}\Big[
&{R\over8\pi}-2(\nabla\phi)^2-e^{-2\alpha_0\phi}F^2\\
&-\tfrac12 e^{4\alpha_1\phi}(\nabla\kappa)^2
-\kappa\,F_{ab}(*F)^{ab}
\Big],
\end{aligned}
\end{equation}
where the first term is the standard Hilbert action and \(\phi\) and \(\kappa\) are the dilaton and axion, respectively.
\(F_{ab}\) is the U(1) field strength, and \((*F)_{ab}\) is
its dual. The parameters $\alpha_0$ and $\alpha_1$ are real, positive constants
that set the coupling strengths and allow us to consider a family of EMDA
theories. Particular choices correspond to particular theories. For
instance, $\alpha_0=\alpha_1=0$ gives Einstein--Maxwell theory minimally
coupled to scalar and axion fields, while
$(\alpha_0,\alpha_1)=(\sqrt{3},0)$ corresponds to Kaluza--Klein.
In this work we take $(\alpha_0,\alpha_1)=(1,1)$, corresponding to a
low-energy limit of heterotic string theory. For these coupling values, the
theory admits the Kerr--Sen family of rotating, charged black hole
solutions. These solutions reduce to Kerr when the electromagnetic charge
vanishes. In contrast to the Kerr--Newman solution of Einstein--Maxwell
theory, the Kerr--Sen geometry is accompanied by nontrivial dilaton and
axion fields whose structure is determined by the black hole charge and
spin. The explicit Kerr--Sen metric and associated fields are summarized in
Appendix~\ref{sec:appx:KerrSenBH}.

\section{Numerical methods}
\label{sec:methods}

With the aim of studying the nonlinear dynamics of EMDA in black hole
collisions, we express EMDA as a Cauchy problem and evolve the
metric using the BSSN formulation~\cite{Shibata1995_BSSN,
Baumgarte1999_BSSN, Baumgarte2010}, supplemented by evolution equations for the electromagnetic, dilaton, and axion fields. We also employ puncture gauge conditions~\cite{Campanelli2006,Baker2006}. The full equations of motion and the $3+1$ decomposition are presented in Appendix~\ref{sec:appx:eom}.  Initial data are prepared for head-on black hole collisions and are described below.

\subsection{Code Description}
\label{sec:code_description}

We perform our black hole merger simulations using \texttt{Dendro-GR}, a well developed numerical relativity framework. \texttt{Dendro-GR} combines an octree-based adaptive
mesh infrastructure with automatic code generation for the evolution
equations~\cite{Fernando2019,Fernando2019b,Fernando2023}. The computational domain is refined dynamically using wavelet adaptive multiresolution (WAMR) in which the evolved fields are represented on an adaptive hierarchy of octree blocks. Wavelet coefficients computed from the evolved variables provide a local estimate of the truncation error 
or unresolved field content~\cite{DeBuhr2018RelativisticHydroWavelets,Anderson2006RelativisticMHDAMR, Fernando2023}.
Regions where these wavelet coefficients exceed a prescribed tolerance are refined, while regions in which wavelet coefficients remain below a coarsening threshold are de-refined. This tends to concentrate resolution near the black holes, in the wave extraction region, and in regions where the electromagnetic, dilaton, or axion fields develop large spatial gradients, while smoother regions are evolved on coarser grids.

Spatial derivatives are computed using sixth-order finite-difference
stencils, and time integration is performed using a fourth-order
Runge--Kutta scheme. We include Kreiss--Oliger dissipation to damp
poorly resolved high-frequency modes and improve numerical stability
\cite{KreissOliger1973}. Additional details of the \texttt{Dendro-GR}
infrastructure, including its WAMR implementation, scalability, and previous applications to binary black hole evolutions, can be 
found in~\cite{Fernando2019,Fernando2019b,Fernando2023,Black2025_ORBIT}.

During the evolutions, we monitor the Hamiltonian and momentum constraints together with the electromagnetic divergence constraints throughout the computational domain. These quantities provide a global check on the consistency of the evolution and are used to assess the efficacy of the WAMR refinement parameters. Representative plots of the constraints and their violation are shown in Appendix~\ref{sec:code_performance}.

We also compute quasilocal quantities on apparent horizons using the
\texttt{BHaHAHA} apparent horizon finder~\cite{Etienne2026BHaHAHA}, which
we have extended to evaluate the horizon mass, spin, and electric charge.
These quantities provide independent diagnostics of remnant relaxation and
of the conservation of the black hole charges during the evolution.

As a check on our overall numerical strategy, we perform a limited type of convergence study by varying both the wavelet
tolerance and maximum refinement level for a short head-on binary evolution.  We confirm that with increased numbers of refinement levels and tighter refinement criteria constraint violations are reduced.  
The details of this study are presented in Appendix~\ref{sec:code_performance}.

\subsection{Initial Data}

Valid initial data must satisfy the relevant constraint equations. In particular, the Hamiltonian, momentum, and Maxwell constraints are 
\begin{align}
R + K^2 - K_{ij}K^{ij} &= 16 \pi \rho, \\
D_j\left(K^{ij} - \gamma^{ij}K\right) &= 8 \pi S^i, \\
%\end{align}
%\begin{equation}
D_i E^i &= \rho_E, \\
\qquad
D_i B^i &= 0,
\end{align}
where \(R\) is the Ricci scalar associated with the spatial metric
\(\gamma_{ij}\), \(K_{ij}\) is the extrinsic curvature, \(K\) is its trace,
and \(D_i\) is the covariant derivative compatible with \(\gamma_{ij}\).
The electric and magnetic fields are \(E^i\) and \(B^i\) with electric charge density, $\rho_E$, while quantities \(\rho\) and \(S^i\) are the matter energy and current densities,
respectively.  The latter, in EMDA, take the form
\begin{align}
%\begin{split}
\rho_E \equiv& \, 2\alpha_0 E^i D_i \phi + e^{2\alpha_0\phi} B^i D_i \kappa \\
\rho \equiv& \, \Pi^2 + (D\phi)^2
+ e^{-2 \alpha_0 \phi}\big( E_i E^i + B_i B^i\big),  \nonumber\\
&+ \tfrac{1}{4} e^{4 \alpha_1 \phi}\big(\Xi^2 + (D\kappa)^2\big), \\
S^i \equiv& 2 \Pi D^i \phi
+ 2 e^{-2 \alpha_0 \phi} \epsilon^{ijk} E_j B_k
+ \tfrac{1}{2} e^{4 \alpha_1 \phi} \Xi D^i \kappa.
\end{align}
where \(\Pi\) and \(\Xi\) denote the momenta conjugate to the
dilaton \(\phi\) and axion \(\kappa\), respectively.

Binary black hole initial data in EMDA have not been extensively studied,
although several methods have been developed for constructing initial data
for charged black holes in related settings
\cite{Alcubierre2009,Mukherjee2022_ConformallyCurvedChargedSpinning,
Zilhao2014,Zilhao2012,Bozzola2019}. In this work, we adapt some of these techniques to the EMDA system by using an elliptic solver based on a modified version of the two-charged-punctures solver introduced 
in~\cite{Bozzola2019}.
Our implementation includes contributions from the dilaton and axion fields. A more detailed discussion of the initial data construction is provided in Appendix~\ref{sec:appx:id}.

To analyze the radiation emitted, we compute the Newman--Penrose scalar
\(\Psi_4\), along with the electromagnetic radiation. We extract their multipolar components at \(r \approx 100M\).  Additional details of the waveform extraction procedure are  given in Appendix~\ref{sec:appx:wave_extraction}.  A summary of the initial data parameters used in our simulations is provided in Appendix~\ref{sec:appx:sim_tab}.

\section{Results}
\label{sec:results}

We now describe the simulations of two broad classes of binary black holes in EMDA.  The first consist of two Kerr--Sen black holes with at least one having nontrivial dilaton and axion fields attached.  This is a nice, even stringent test of the persistence of these scalar fields through merger and the settling down of the final object to a new and different Kerr--Sen black hole.  The second is the simulation of two Kerr--Newman black holes with perturbative dilaton and axion fields in the vicinity.  This latter case is to test whether the merging black holes can acquire these scalar fields and whether the merger then results in a final Kerr--Sen remnant.

% ---------------------------------------------------------
% Waveforms (stacked subfigures)
% ---------------------------------------------------------
\begin{figure}[!t]
  \centering

  \subfloat[Real part of the $(2,0)$ mode of $\Psi_4$ extracted at $R_{\mathrm{ex}} = 100$.%
    \label{fig:gw_l2m0}%
  ]{%
    \includegraphics[width=\linewidth]{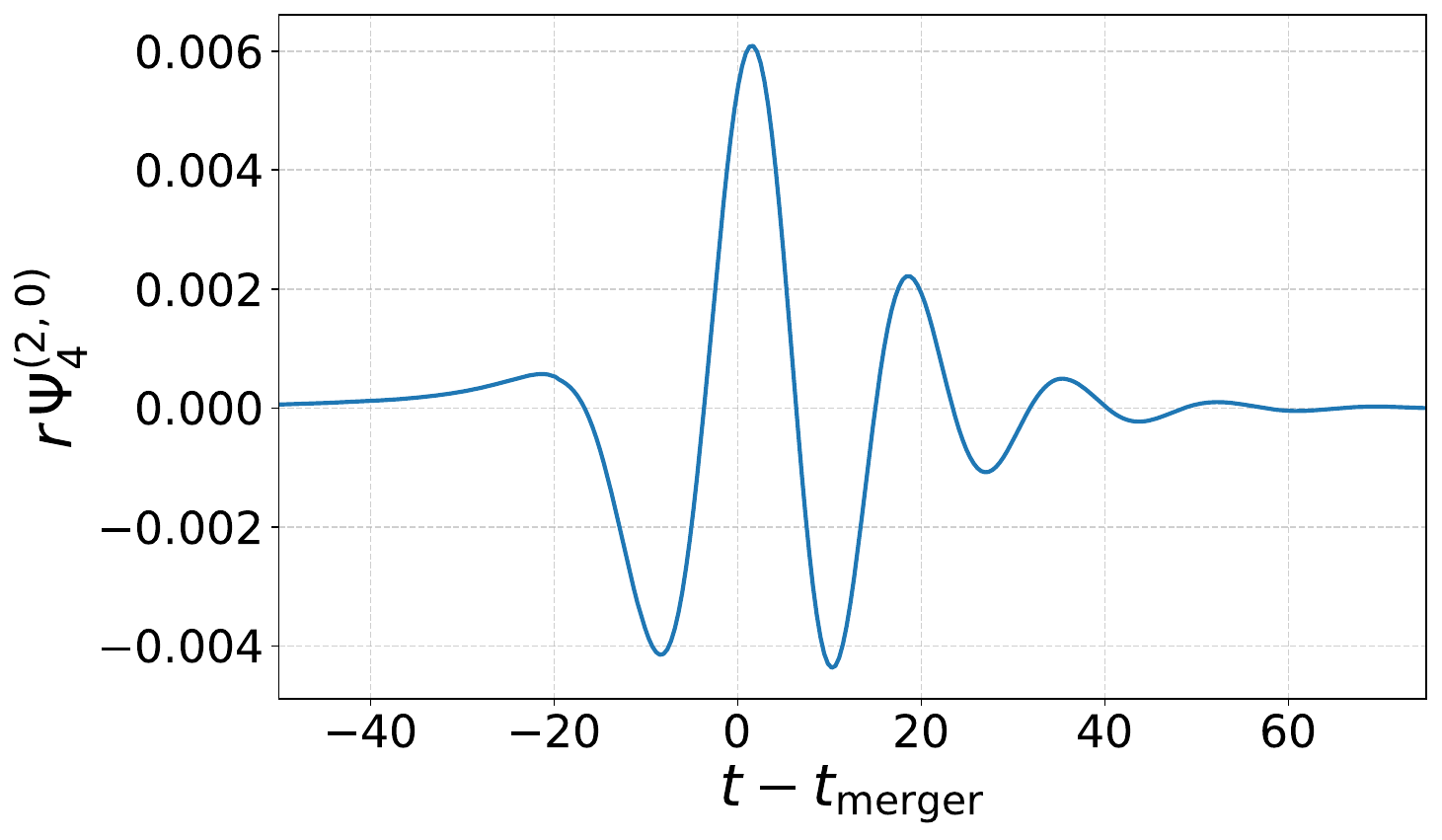}%
  }\\[0.6em]

  \subfloat[Real part of the $(2,0)$ mode of $\Phi_1$ extracted at $R_{\mathrm{ex}} = 100$.%
    \label{fig:phi2_l2m0}%
  ]{%
    \includegraphics[width=\linewidth]{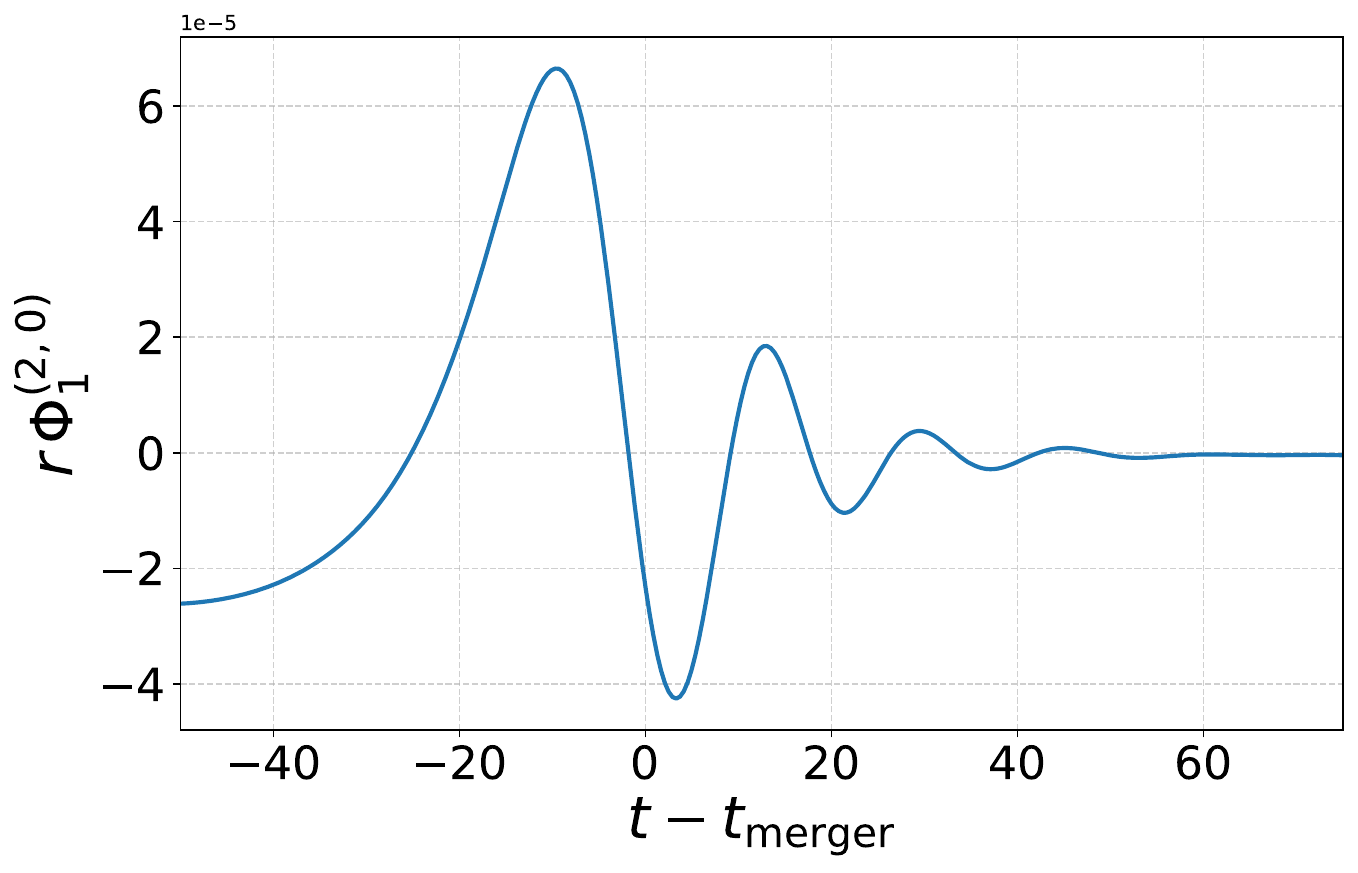}%
  }\\[0.6em]

  \subfloat[Real part of the $(2,0)$ mode of $\Phi_2$ extracted at $R_{\mathrm{ex}} = 100$.%
    \label{fig:phi1_l2m0}%
  ]{%
    \includegraphics[width=\linewidth]{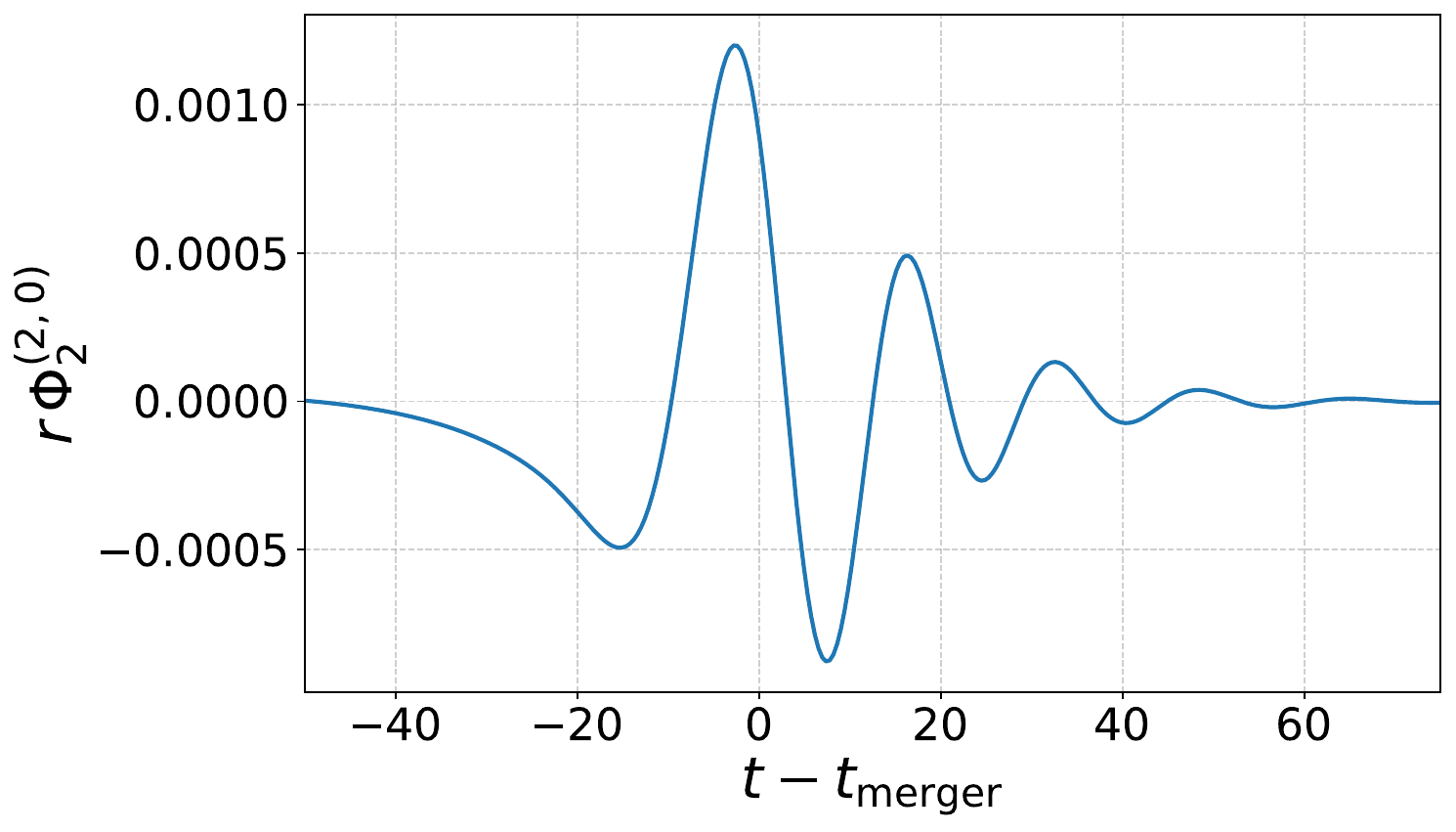}%
  }

\caption{The real portions of the $(2,0)$ modes of the gravitational radiation $\Psi_4$, electromagnetic radiation $\Phi_1$, and $\Phi_2$, extracted at $R_{\mathrm{ex}} = 100M$ for a binary with charge-to-mass ratio $q_i/m_i = 0.1$, dimensionless spin $\chi = 0.4$, and an initial coordinate separation of $16M$.}
  \label{fig:waveforms_l2m0}
\end{figure}

\subsection{Initially Scalarized Black Holes}
We perform simulations across a range of binary Kerr--Sen black hole configurations in the initial data space.  For the current work, we focus on equal-mass binaries with low to moderate charge and low spin.  In particular, the total mass of the system is normalized to $M=1$.  The charge to mass ratio on each black hole ranges between $0.0625 \leq q_i/m_i \leq 0.5$ where $m_i$ and $q_i$ are the initial masses and charges of the individual black holes.  For spinning cases, the spins of both black holes are usually taken to be equal and aligned.  The spin magnitude of each is taken to lie in the range $0.05 \leq \chi \leq 0.4$. A more complete summary of the parameter values used in our tests is provided in two tables as Appendix~\ref{sec:appx:id}. 

\begin{figure}[!t]
  \centering
  \includegraphics[width=\linewidth]{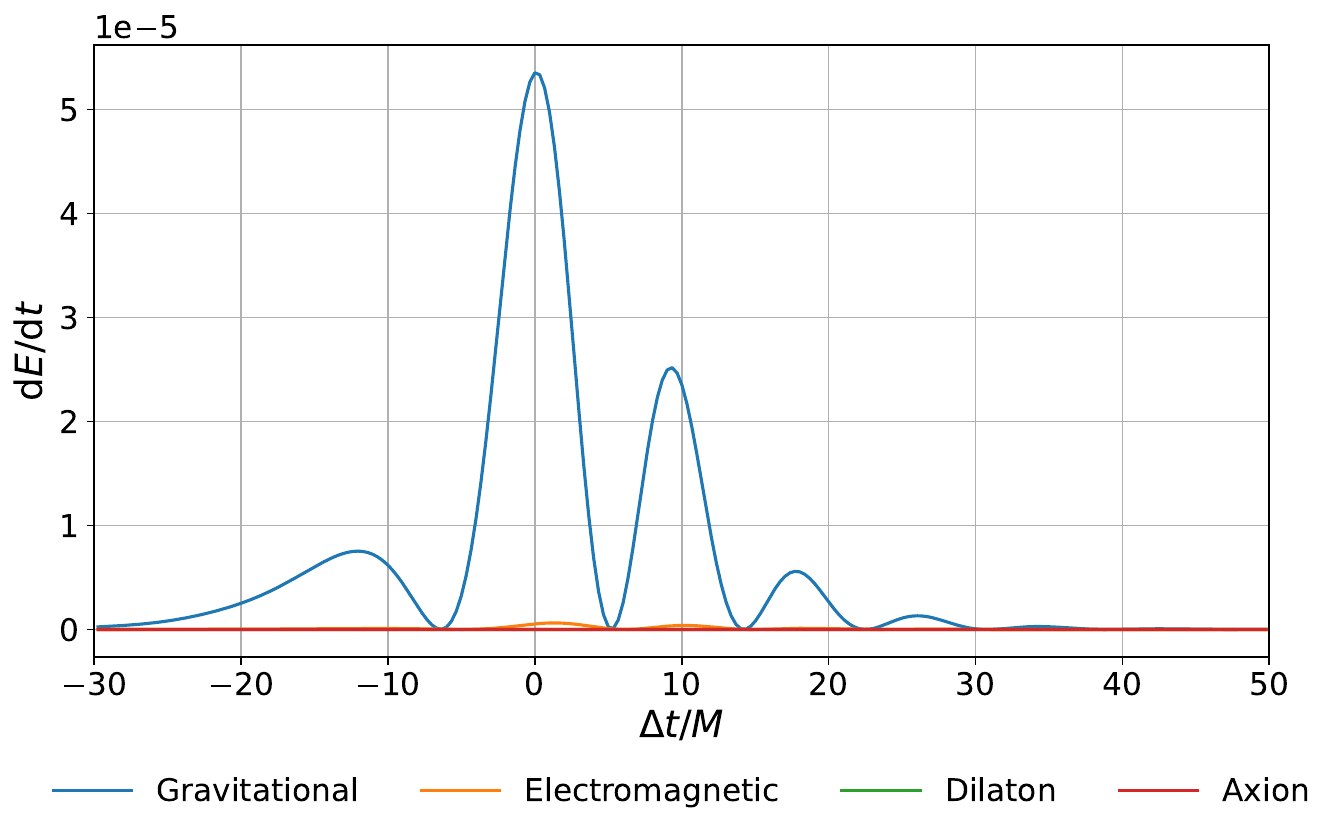}
  \caption{Plot of the radiated energy for the case $\chi = 0.4$ and $q_i/m_i = 0.1$.  The curves have been aligned such that $t = 0$ corresponds to the global maximum. On this plot the axion and dilaton radiation energy remain at least 3 orders of magnitude smaller than the GW or EM contributions}
  \label{fig:flux_chi02_Q05}
\end{figure}

\begin{figure}[!t]
  \centering

  \subfloat[Axion\label{fig:axion}]{
    \includegraphics[width=0.95\linewidth]{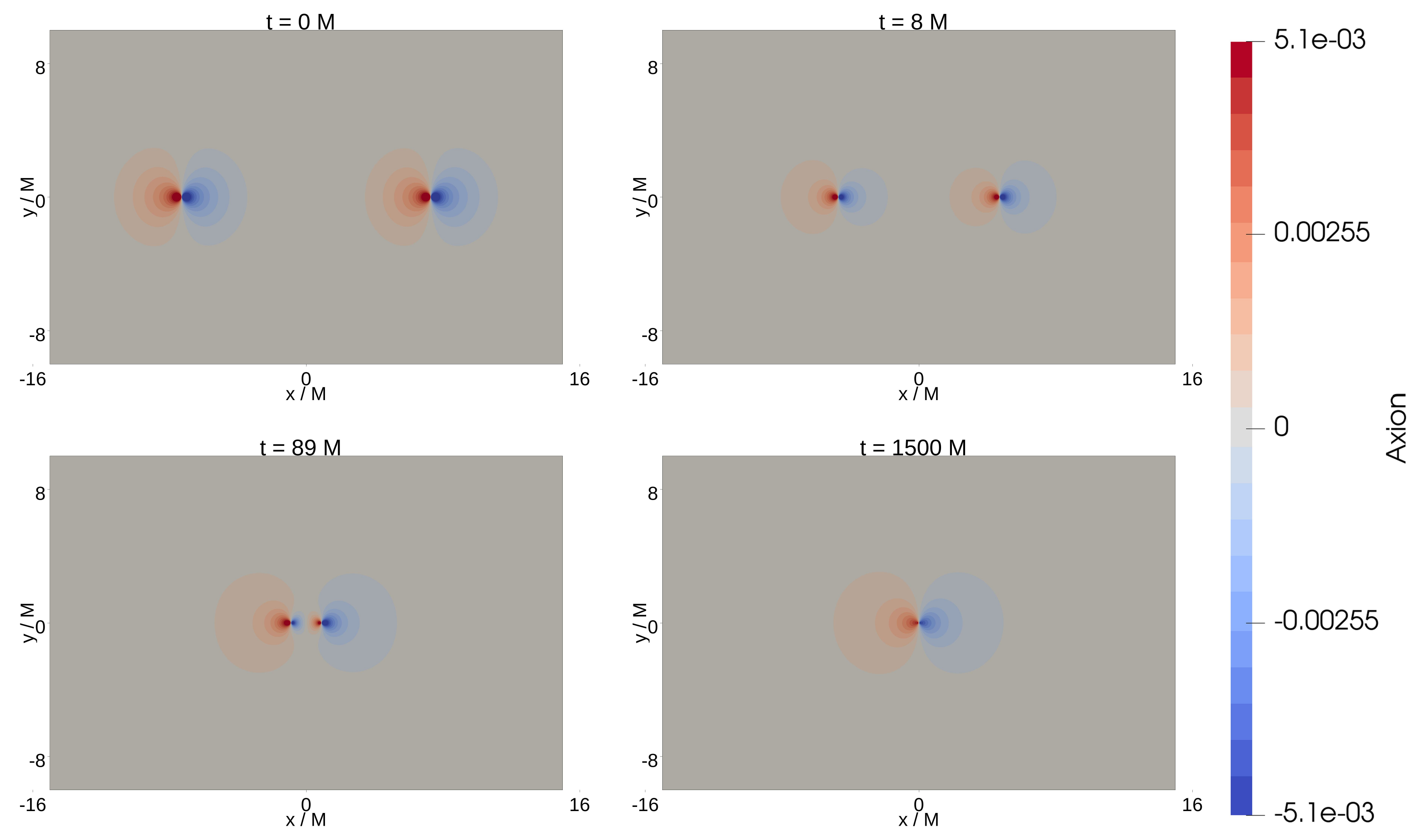}
  }\\[1em]

  \subfloat[Dilaton\label{fig:dilaton}]{
    \includegraphics[width=0.95\linewidth]{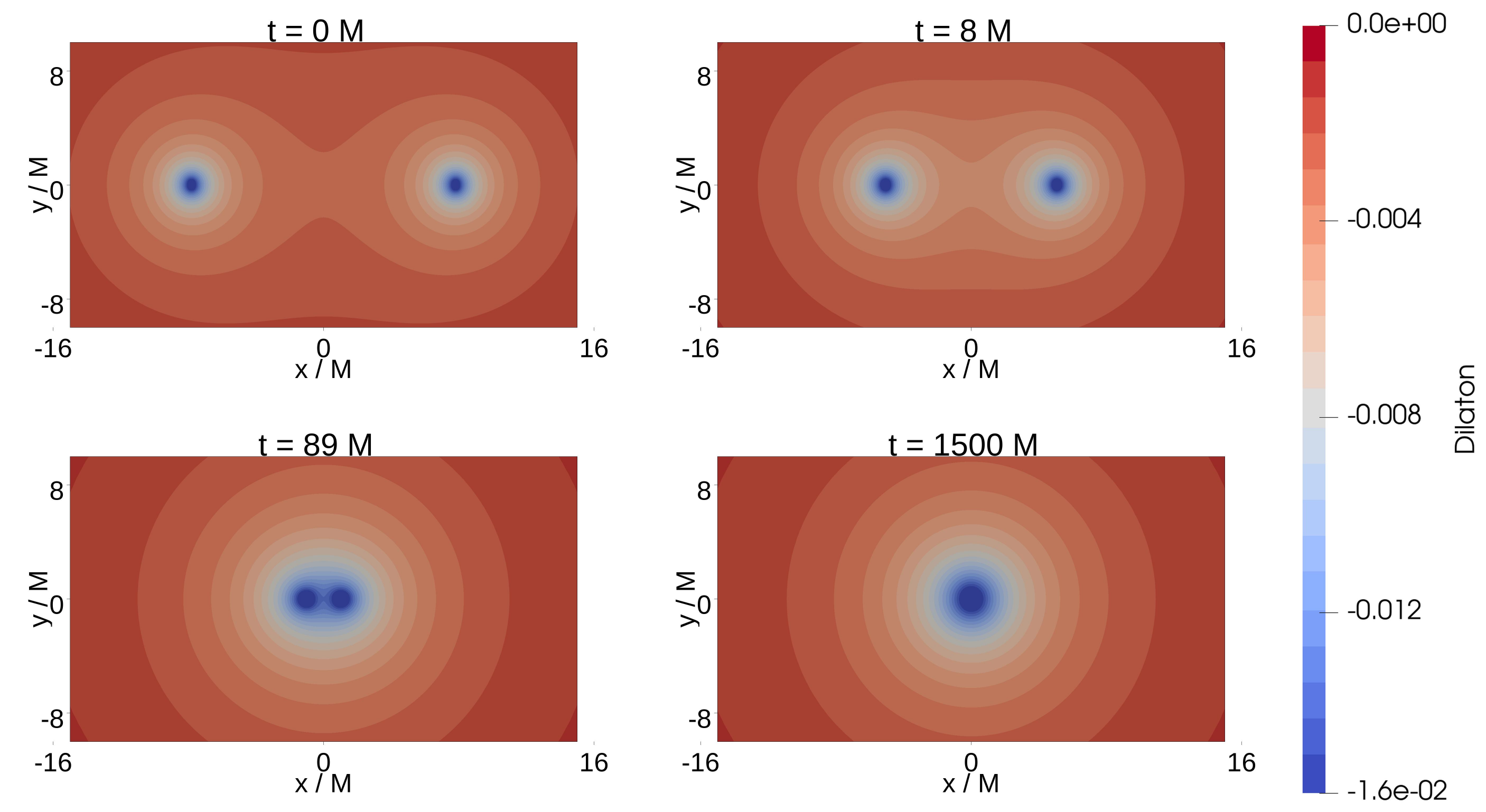}
  }

  \caption{The scalar fields on the $z=0$ plane for the case of $\chi = 0.4$ and $q_i/m_i = 0.1$ Kerr--Sen BHs. We show the amplitudes of the axion, $\kappa$ (top four panels), and dilaton, $\phi$ (bottom four panels), fields at the beginning of the evolution (top left), a short time before merger (top right), just after merger (bottom left), and some time after merger (bottom right).}
  \label{fig:scalar_fields}
\end{figure}

Figure~\ref{fig:waveforms_l2m0} shows the waveforms from one such representative head-on collision of two identical Kerr--Sen black holes.  For this system, both BHs have $\chi = 0.4$ and $q/m = 0.1$, are initially separated by $16M$ and fall together from rest. 
The waveforms are of the gravitational radiation, $\Psi_4$ and electromagnetic radiation,$\Phi_1$ and $\Phi_2$.  
The radiated energy as a function of time is shown in Figure~\ref{fig:flux_chi02_Q05}.  
In this case, the GW is clearly the dominant form of energy released followed by the electromagnetic radiation. The scalar radiation is much smaller than either the EM or GW radiation. We conclude that there is very little energy radiated in the scalar field channels through the evolution.  In all similar experiments, we find that the scalar fields on the black holes persist and do not  radiate away through or following the merger.  Nontrivial dilaton and axion fields are clearly present and remain attached to the black hole after the merger, as illustrated in Figure~\ref{fig:scalar_fields}.

\begin{figure}[tbp]
  \centering
  \includegraphics[width=0.92\columnwidth]{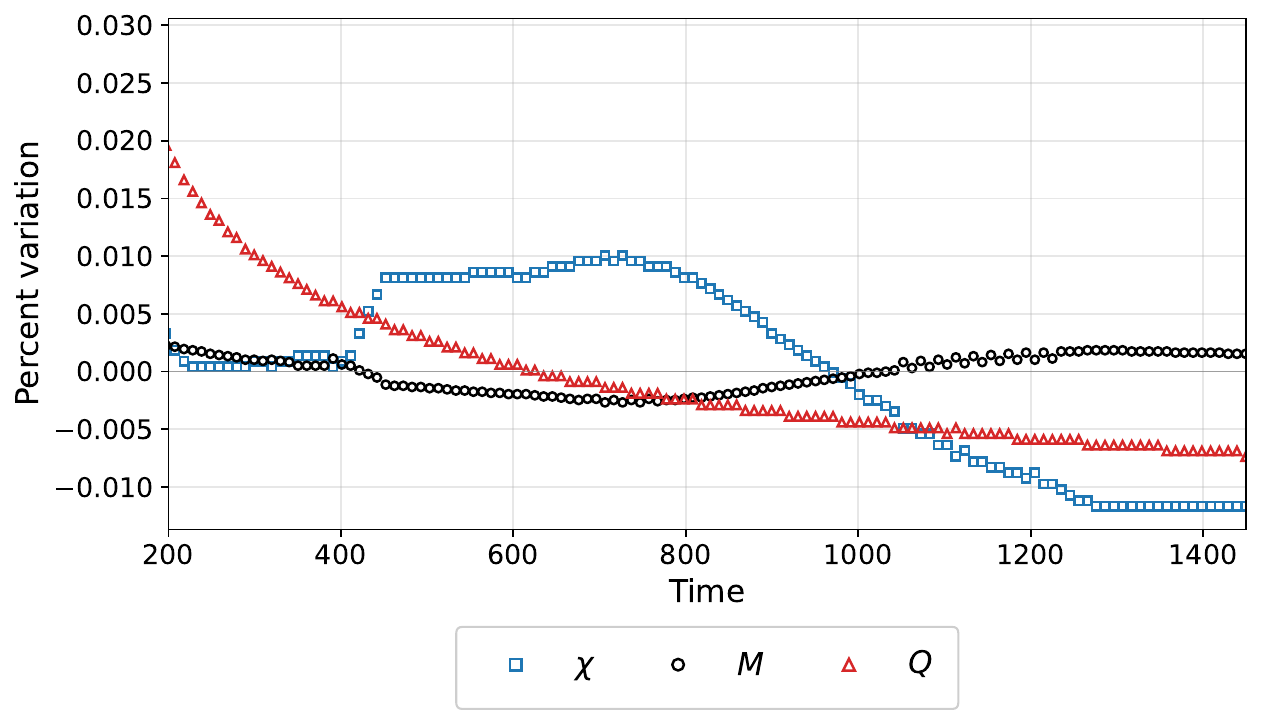}
  \caption{
  Percent deviation of the remnant black hole's horizon mass $M$, electric
  charge $Q$, and angular momentum $\chi$ from their calculated average values: 
  $\langle M \rangle = 0.9682$, $\langle Q \rangle = 0.2001$, and
  $\langle \chi \rangle = 0.1961$.  These quantities are
  measured on the apparent horizon of the remnant black hole for an initially
  spinning and charged binary black hole configuration with individual
  charge-to-mass ratios $q_i/m_i = 0.1$, angular momenta $\chi = 0.4$, and total mass $M=1$.
  }
  \label{fig:ql_m_q_Jz}
\end{figure}

We monitor quasilocal quantities on the apparent horizon to verify that the remnant black hole is stable.  In particular, we find that the horizon mass, dimensionless spin, and electric charge settle down to approximately constant values.  Figure~\ref{fig:ql_m_q_Jz} shows the percent variation in these quantities after the final black hole has settled into a stationary configuration.  In general, these quantities exhibit only small
late-time variations, indicating that the remnant has settled to a stable
final configuration. Indeed, with these calculated black hole values, we are able to confirm that on comparison with the value of the dilaton on the horizon, the resulting black hole is another, larger mass Kerr--Sen black hole with dilaton and axion fields.

% ---------------------------------------------------------
% Radiated energy / flux figure
% ---------------------------------------------------------

% -------- Scalar field profiles (stacked top-to-bottom) --------

% ---------------------------------------------------------
% Remnant black hole quasilocal quantities
% ---------------------------------------------------------

\subsection{Initially Unscalarized Black Holes}

To further assess the stability of scalar hair, we conducted a series of tests initialized with a Gaussian distribution of axion and dilaton fields centered at the origin, with amplitudes ranging from $10^{-7}$ up to almost $10^{-2}$. 
Despite not strictly satisfying the constraints, these superposed perturbative scalar fields propagate outward and enable us to examine whether initially unscalarized black holes can dynamically acquire and retain scalar fields. Our results confirm that they do: a Kerr–Newman black hole in EMDA rapidly scalarizes, developing both dilaton and axion hair. In addition, the change to the constraint violations both from the initial perturbations as well as through the subsequent evolution are small enough to leave us confident in our results.  

\begin{figure}[!t]
  \centering

  \subfloat[Axion\label{fig:axion}]{
    \includegraphics[width=0.95\linewidth]{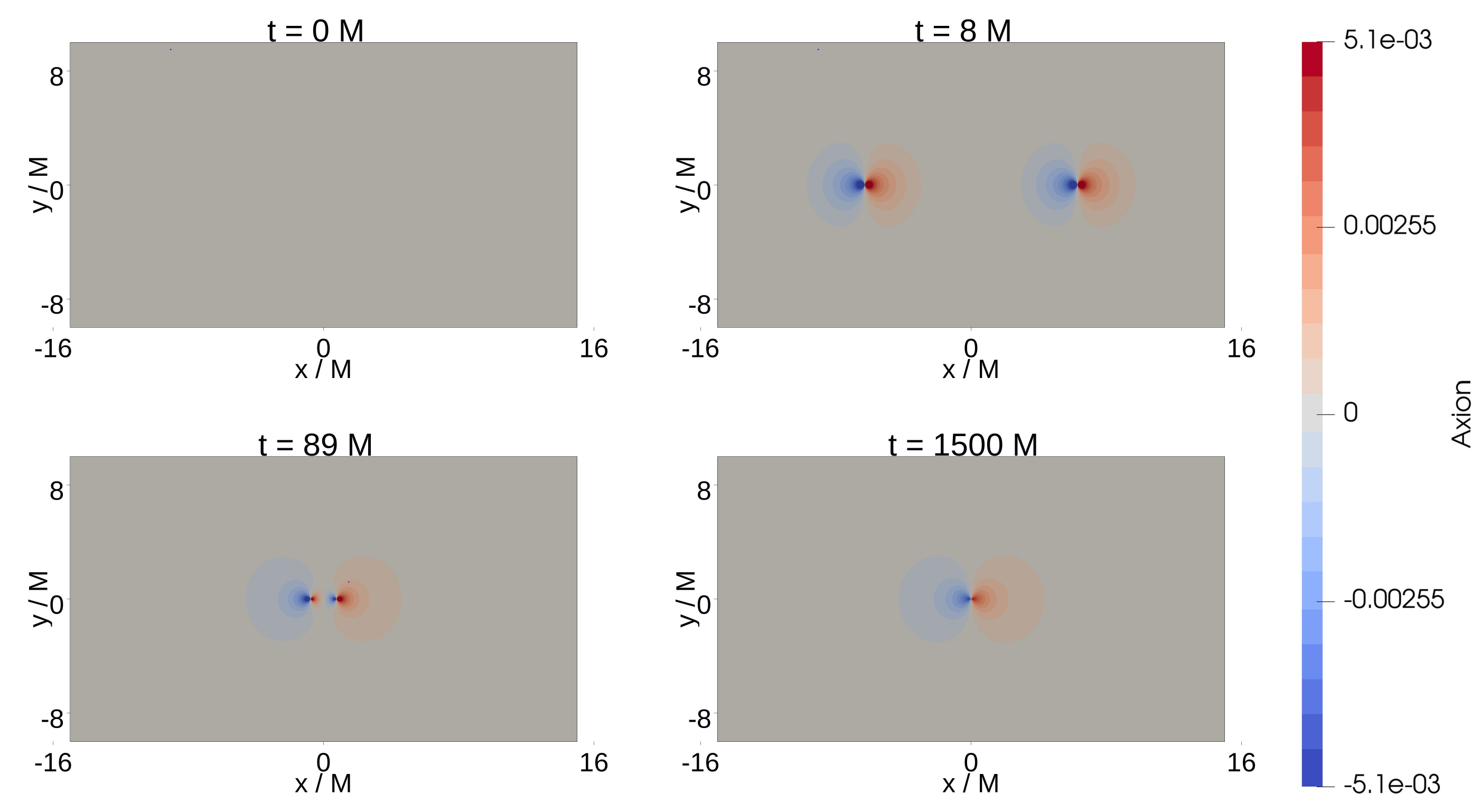}
  }\\[1em]

  \subfloat[Dilaton\label{fig:dilaton}]{
    \includegraphics[width=0.95\linewidth]{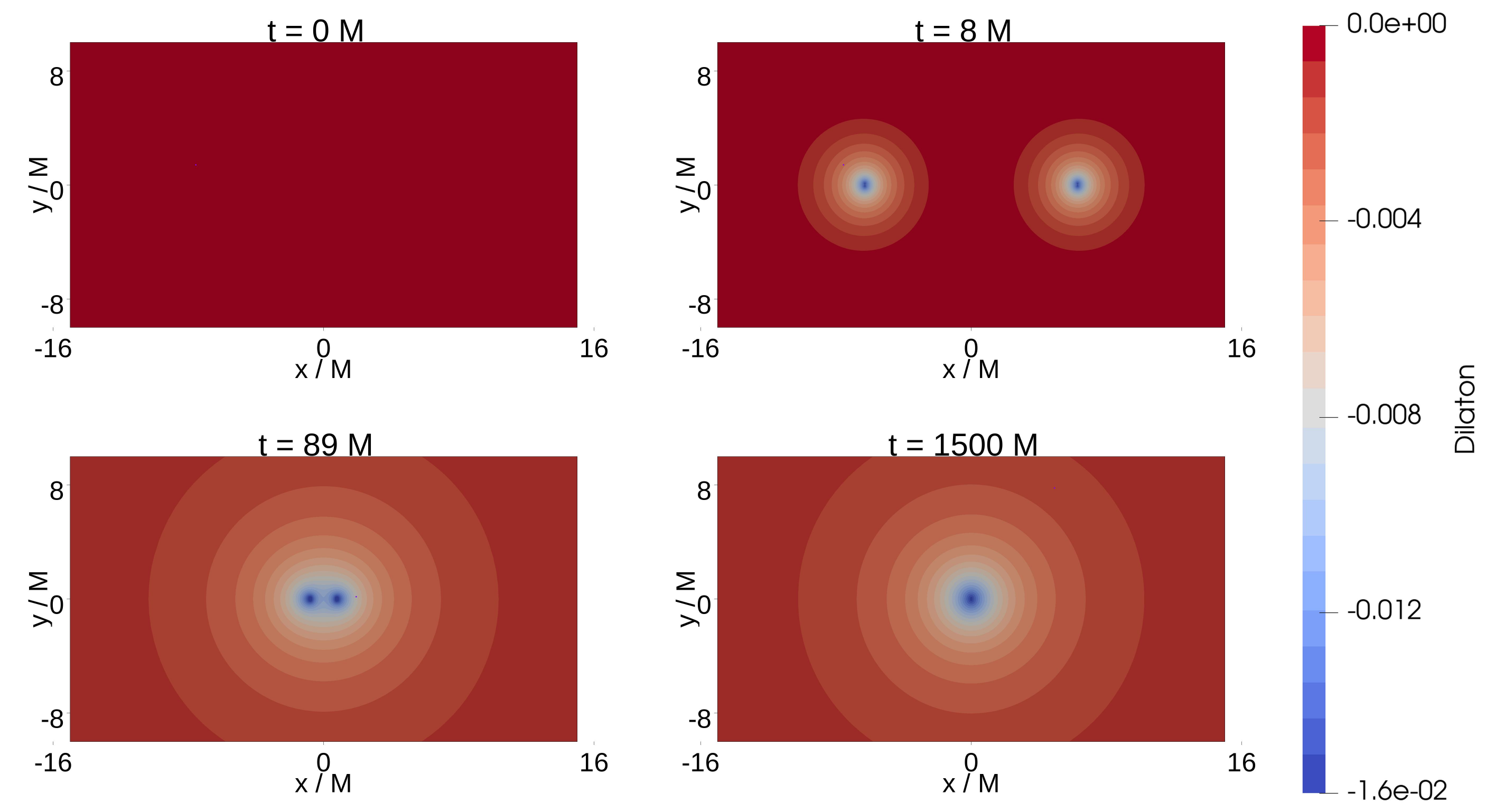}
  }

  \caption{The scalar fields on the $z=0$ plane for the case of Kerr--Newman BHs initially set to $\chi = 0.4$ and $q_i/m_i = 0.1$.  Neither begin by carrying dilaton or axion fields.  
  Shown are the amplitudes of the scalar fields $\phi$ and $\kappa$ at various stages of the evolution: 
  the initial stage (top left), where both fields begin at zero; 
  a short time later as the fields start to develop (top right); 
  just before merger (bottom left); and some time after merger (bottom right).}
  \label{fig:unscalarized}
\end{figure}

Even in the absence of any initial scalar fields in the background, the system evolves to generate nontrivial scalar structure. This demonstrates that scalarization occurs dynamically and does not require pre-existing scalar hair. The scalarization occurs quickly with initially Kerr--Newman BHs acquiring Kerr--Sen aspects, i.e. scalar hair prior to merger.  Mass and electromagnetic charge get radiated with a very small amount of scalar hair, but the final object is a Kerr--Sen BH.  
%The conserved quantities, including the total mass and charge, remain effectively constant throughout the evolution. Variations in the conserved quantities over the course of the evolution in both mass and charge are below \(0.2\,\%\). This suggests that these quantities remain stable and conserved over time. 
Figure~\ref{fig:unscalarized} illustrates the evolution of the scalar fields for the case without initial scalar hair on the BHs.

% -------- Scalar field profiles (stacked top-to-bottom) --------

\subsection{Constraint and refinement tests}
\label{sec:numerical_diagnostics}

As a consistency check on the evolutions, we monitor the constraint
violations throughout the simulations. Representative full evolution
constraint plots are shown in Appendix~\ref{sec:code_performance}. These
diagnostics show that the Hamiltonian, momentum, and electromagnetic
constraint violations remain controlled over the time scales considered
here, including through the merger and ringdown.

We also perform a limited refinement study for a short head-on binary
evolution.  In this study, we vary both the wavelet tolerance and the maximum number of refinement levels.  Both are WAMR related parameters that help dictate the amount and extent of refinement.  The wavelet tolerance is varied between $3\times10^{-5}$ and $3\times10^{-6}$,  
and we consider a maximum number of refinement levels between 14 and 17, corresponding to grid spacings between $h_{14}=0.0325$ and $h_{17}=0.00407$.  
Additional details are given in
Appendix~\ref{sec:code_performance}. Tightening the refinement criteria
reduces the volume weighted Hamiltonian and momentum constraint violations,
consistent with the expected behavior of the WAMR refinement scheme. Taken
together, the full evolution constraint monitoring and refinement tests
indicate that the simulations remain well behaved over the time scales
considered here.

\section{Conclusion}
\label{sec:conclusion}
We have simulated a range of different parameters for the head-on collision of Kerr--Sen black holes. We have primarily been concerned with the persistence of scalar hair through this highly nonlinear, dynamical process.  
In summary, our simulations show that the scalar fields do not radiate away so long as there is a nonzero charge (and spin for the axion). We have found that Kerr--Sen black holes are stable. In particular, we find that if the remnant black hole has a nonzero charge, the dilaton field persists. If it has both nonzero charge and spin, the axion field also persists after merger. These results suggest that Kerr--Sen black holes are stable under the conditions and time frames we considered.  

A natural follow-up study would be the analysis of waveforms produced by inspiraling black holes, an area where the available results remain relatively limited. Further, while we have assumed a particular member of the family of EMDA theories in this work, namely a low-energy limit of heterotic string theory, there is a much larger theory space parametrized by the dilaton and axion couplings, $\alpha_0$ and $\alpha_1$, that remains largely unexplored. A more detailed investigation of this theory space could be intriguing; possibly giving rise to broader classes of stable black holes as well as detectable departures from the predicted general relativistic gravitational wave signal.

\begin{acknowledgments}
We would like to thank several individuals for their help and contributions to this work: Gabriele Bozzola for several useful discussions and for sharing his two-charged-puncture code with us; Zach Etienne for help with \texttt{BHaHAHA}; Milinda Fernando for his extensive insights on the \texttt{Dendro} framework; William Black for many discussions and improvements to \texttt{Dendro-GR}.  In addition, thanks to Luis Lehner, Steven Liebling and Carlos Palenzuela for early work on EMD. This research was supported by NSF grant PHY-2207615 (BYU). HL is supported by the LANL ASC program. LANL is operated by Triad National Security, LLC, for the National Nuclear Security Administration of the U.S. DOE (Contract No.~89233218CNA000001). This work is authorized for unlimited release under LA-UR-25-31225.

\medskip
\noindent\textit{Artificial intelligence assistance.}
The authors acknowledge the limited use of artificial intelligence tools,
including ChatGPT and Gemini, during the preparation of this manuscript.
These tools were used for spelling and grammar checks, LaTeX editing and
debugging, and improving the clarity and readability of some text. They
were also used to assist with code debugging and code development. The
authors reviewed and verified all AI-assisted material and take full
responsibility for the scientific content, calculations, code,
interpretations, and conclusions presented in this work.
\end{acknowledgments}

\appendix
\section{Kerr--Sen Metric}
\label{sec:appx:KerrSenBH}
We present the Kerr--Sen solution in the Einstein frame, written in Boyer--Lindquist type coordinates. The metric takes the form

\begin{align}
ds^2 &= -\left(1-\frac{2mr\cosh^2\alpha}{\rho^2}\right)dt^2 + \rho^2\left(\frac{dr^2}{\Delta}+d\theta^2\right) \notag \\
&\quad+\frac{\Sigma}{\rho^2}\sin^2\theta\, d\varphi^2
-\frac{4amr\cosh^2\alpha}{\rho^2}\sin^2\theta\, dt\, d\varphi
\end{align}
where the metric functions are defined by
\begin{align}
\Delta &= r^2 - 2mr + a^2, \\
\rho^2 &= r^2 + a^2\cos^2\theta + 2mr\sinh^2\alpha, \\
\Sigma &= \left(r^2 + a^2 + 2mr\sinh^2\alpha\right)^2
         - \Delta a^2 \sin^2\theta.
\end{align}
The associated gauge, dilaton, and axion fields are given by
\begin{align}
A_a dx^a &= \frac{mr \sinh 2\alpha}{\sqrt{2}\,\rho^2}
\left(dt - a \sin^2\theta \, d\varphi \right), \\
\label{Kerr-Sen_dilaton}
e^{-2\phi} &= \frac{\rho^2}{r^2 + a^2 \cos^2\theta}, \\ 
\kappa &= -2ma \sinh^2\alpha \, \frac{\cos\theta}{r^2 + a^2 \cos^2\theta}.
\end{align}

The parameters $m$, $\alpha$, and $a$ arise as integration constants. The physical mass $M$, electric charge $Q$, angular momentum $J$, and magnetic dipole moment $\mu$ are related to these parameters via
\begin{align}
M &= m \cosh^2\alpha, \\
Q &= \frac{m}{\sqrt{2}} \sinh 2\alpha, \\
J &= aM, \\
\mu &= aQ.
\end{align}

With respect to the Boyer--Lindquist radial coordinate, $r$, the event horizon is located at the familiar 
\begin{equation}
r_H = m + \sqrt{m^2 - a^2},
\end{equation}
and the condition for the existence of a nonextremal black hole %(i.e., the absence of a naked singularity) 
is
\begin{equation}
m^2 > a^2 \quad \Leftrightarrow \quad M^2 > |J| + \frac{Q^2}{2}.
\end{equation}
The inverse relations are
\begin{align}
m &= M - \frac{Q^2}{2M}, \\
\sinh 2\alpha &= \frac{2\sqrt{2}\,QM}{2M^2 - Q^2}, \\
a &= \frac{J}{M}.
\end{align}

\section{Equations of Motion}
\label{sec:appx:eom}

The action for EMDA in the Einstein frame is given as:
\begin{equation}
\begin{aligned}
S=\int \mathrm{d}^4 x\,\sqrt{-g}\Big[
&{R\over8\pi}-2(\nabla\phi)^2-e^{-2\alpha_0\phi}F^2\\
&-\tfrac12 e^{4\alpha_1\phi}(\nabla\kappa)^2
-\kappa\,F_{ab}(*F)^{ab}
\Big]
\end{aligned}
\end{equation}
where $\phi$ and $\kappa$ are the dilaton and axion, respectively.
The tensor $F_{ab}$ is the U(1) field strength, and $(*F)_{ab}$ is its
dual. In this work we set the coupling constants to
$\alpha_0=\alpha_1=1$, corresponding to a low-energy limit of heterotic
string theory.

The Einstein equations are:
\begin{equation}
\begin{split}
%R_{ab} - \tfrac{1}{2} g_{ab} R
G_{ab} = 8\pi G\bigg[&
 \, 2\left(
    \nabla_a \phi \nabla_b \phi
    - \tfrac{1}{2} g_{ab}(\nabla \phi)^2
\right) \\
&+ 2 e^{-2\alpha_0\phi}
\left(
    F_{ac}F_b{}^c
    - \tfrac{1}{4} g_{ab}F^2
\right) \\
&+ \tfrac{1}{2} e^{4\alpha_1\phi}
\left(
    \nabla_a\kappa \nabla_b\kappa
    - \tfrac{1}{2} g_{ab}(\nabla \kappa)^2
\right)
\bigg] 
\end{split}
\label{eq:Einstein}
\end{equation}
while the matter equations can be written 
\begin{align}
0 &= \nabla^a \nabla_a \phi
   + \tfrac{1}{2}\alpha_0 e^{-2\alpha_0 \phi} F^2
   - \tfrac{1}{2}\alpha_1 e^{4\alpha_1 \phi} (\nabla \kappa)^2, \label{eq:phi} \\[1ex]
0 &= \nabla^a \!\left( e^{4\alpha_1 \phi} \nabla_a \kappa \right)
   - F_{ab} (\ast F)^{ab}, \label{eq:kappa} \\[1ex]
0 &= \nabla_a \!\left( e^{-2\alpha_0 \phi} F^{ab} \right)
   + (\ast F)^{ab} \nabla_a \kappa. \label{eq:Maxwell}
\end{align}

We include a form of hyperbolic constraint damping into the Maxwell equations so that violations propagate away at the speed of light. In particular, we introduce two scalar fields $\Phi$ and $\Psi$ into \ref{eq:Maxwell} and the identity $\nabla_a(*F)^{ab}=0$ as follows:
\begin{align}
\nabla_{a}\left(F^{ab} + g^{ab}\Psi \right) 
  &= \eta_{1} n^{a} \Psi 
     + 2 \alpha_{0} \nabla_{a}\phi\, F^{ab} \nonumber \\
  &\quad - e^{2\alpha_0 \phi} (*F)^{ab}\nabla_a \kappa, \\[6pt]
\nabla_{a}\!\left( (\ast F)^{ab} + g^{ab}\Phi \right) 
  &= \eta_{2} n^{a} \Phi .
\end{align}
where $\eta_1$ and $\eta_2$ are taken as positive constants and serve as tunable parameters to effect the damping of the elctromagnetic constraints.  Note that on the right hand sides, we have also included $n^a$, the usual unit timelike normal to the spatial hypersurfaces in anticipation of performing the usual $3+1$ split of spacetime.  It is worth noting that these damped versions of Maxwell are now no longer strictly 4-covariant.  

We introduce the usual $3+1$ decomposition of spacetime by defining the line element as 
\begin{align}
{\rm d}s^2 = -\alpha^2 {\rm d} t^2 + \gamma_{ij} ({\rm d}x^i + \beta^i {\rm d}t) ({\rm d}x^j + \beta^j {\rm d}t) 
\end{align}
with lapse, $\alpha$, shift, $\beta^i$, 3-metric, $\gamma_{ij}$, unit timelike normal $n_a = (-\alpha,0,0,0)$ and extrinsic curvature $K_{ab} = - \gamma_a{}^c \gamma_b{}^d\nabla_c n_d$.  

Using the BSSN formulation, we solve the standard 
BSSN evolution equations for the usual $\chi, {\tilde \gamma}_{ij}, K, {\tilde A}_{ij}$ and ${\tilde \Gamma}^i$ variables with EMDA matter terms.\footnote{Note that in this Appendix only, $\chi$ represents the conformal factor and not the dimensionless spin parameter.}  These equations take the form 

\begin{align}
\partial_0\tilde{\gamma}_{ij} &= -2\alpha\,\tilde{A}_{ij},\\[4pt]
\partial_0 \chi &= \tfrac{2}{3}\alpha\,\chi\,K,\\[4pt]
\partial_0 K &= [\ldots] + 4\pi\alpha\,(\rho + T),\\[4pt]
\partial_0\tilde{A}_{ij} &= [\ldots] - 8\pi\chi\alpha\left( T_{ij} - \tfrac{1}{3}\,\gamma_{ij} T_k{}^k\right),\\[4pt]
\partial_0 \tilde{\Gamma}^i &= [\ldots] - 16\pi\alpha\,\chi^{-1} S^{\,i},
%\tilde{\Gamma}^i &= \tilde{\gamma}^{jk}\tilde{\Gamma}^{\,i}{}_{jk}
\end{align}
where $\partial_0 \equiv \partial_t - {\cal L}_{\beta}$ with ${\cal L}_{\beta}$ the Lie derivative along the shift and the $[\ldots]$ are a stand-in for the relevant right hand sides which can be found in standard references, e.g.~\cite{Baumgarte2010} 

The quantities $S^i$, $T_{ij}$, and $\rho$ are defined as
\begin{equation}
\begin{aligned}
S^i
&\equiv -\gamma^{ia} T_{ab} n^b  \\
&=   2\Pi D^i\phi
   + 2e^{-2\alpha_0\phi}\epsilon^{ijk}E_jB_k  %\\
%&\qquad\qquad
   + \tfrac{1}{2}e^{4\alpha_1\phi}\Xi D^i\kappa ,
\label{eq:Si}
\end{aligned}
\end{equation}
\begin{equation}
\begin{aligned}
T_{ij}
&\equiv T_{ab} \gamma^a{}_i \gamma^b{}_j  \\
&= 2\Bigl[
      D_i\phi D_j\phi
      - \tfrac{1}{2}\tilde{\gamma}_{ij}\bigl((D\phi)^2-\Pi^2\bigr)
   \Bigr]  \\
&\quad\quad
- 2e^{-2\alpha_0\phi}\Bigl( 
       E_iE_j + B_iB_j  \\
&\qquad\qquad\qquad\quad
      - \tfrac{1}{2}\tilde{\gamma}_{ij}
        \bigl(E_kE^k + B_kB^k\bigr)
   \Bigr)  \\
&\quad\,
+ \tfrac{1}{2}e^{4\alpha_1\phi}\Bigl[
      D_i\kappa D_j\kappa
      - \tfrac{1}{2}\tilde{\gamma}_{ij}\bigl((D\kappa)^2-\Xi^2\bigr)
   \Bigr] .
\label{eq:Tij}
\end{aligned}
\end{equation}

\begin{equation}
\begin{split}
    \rho \equiv\;& \Pi^2 + (D\phi)^2 
    + e^{-2 \alpha_0 \phi}\big( E_a  E^a +  B_a  B^a\big) \\
    &+ \tfrac{1}{4} e^{4 \alpha_1 \phi}\big(\Xi^2 + (D\kappa)^2\big).
\end{split}
\end{equation}

We use standard puncture gauge conditions for the lapse and the shift:
\begin{align}
\partial_t\alpha &=\beta^i \partial_i - 2\alpha K , \\[4pt]
\partial_t \beta^i &= \beta^j \partial_j \beta^i + \tfrac{3}{4} B^i , \\[4pt] 
\partial_t B^i &= \beta^j \partial_j B^i + \partial_t {\tilde\Gamma}^i - \beta^j \partial_j {\tilde \Gamma}^i - \eta B^i .
\end{align}
We also employ a slow-start lapse condition~\cite{Etienne2024} to reduce transients at early times.  This amounts to adding the term 
\begin{equation}
- \chi^{1/2} \bigl(\varpi \, e^{-t^2/2\sigma^2}\bigr)\,(\alpha - \chi^{1/2}) \\[4pt]
\end{equation}
to the equation for the lapse.  Following~\cite{Etienne2024}, we take the constants $\varpi=3/5$ and $\sigma=20$.

For the matter equations we start with the dilaton and axion:
\begin{align}
    \partial_0 \phi &= - \alpha \Pi , \\
    \partial_0 \kappa &= - \Xi 
\end{align}
\begin{equation}
\begin{aligned}
\partial_0 \Pi
=& \, \alpha K \Pi
- \alpha \chi\, \tilde{\gamma}^{ij} \partial_i \partial_j \phi
- \chi\, \tilde{\gamma}^{ij} \partial_i \alpha\, \partial_j \phi  \\
&+ \alpha \chi\, \tilde{\Gamma}^i \partial_i \phi
+ \frac{\alpha}{2}\tilde{\gamma}^{ij}
  \partial_i \chi\, \partial_j \phi \\
&+ \frac{\alpha\alpha_0}{\chi}
  e^{-2\alpha_0\phi}
  \tilde{\gamma}_{ij}
  \left(E^iE^j - B^iB^j\right)  \\
&+ \frac{\alpha\alpha_1}{2}
  e^{4\alpha_1\phi}
  \left(
    \chi\,\tilde{\gamma}^{ij}
    \partial_i\kappa\,\partial_j\kappa
    - \Xi^2
  \right).
\end{aligned}
\end{equation}

\begin{equation}
\begin{aligned}
\partial_0 \Xi
&= \alpha K \Xi
\\[3pt]
&\quad
- \alpha \chi\, \tilde{\gamma}^{ij} \partial_i \partial_j \kappa
- \chi\, \tilde{\gamma}^{ij}\, \partial_i \alpha\, \partial_j \kappa
\\[3pt]
&\quad
+ \alpha \chi\, \tilde{\Gamma}^i \partial_i \kappa
+ \frac{\alpha}{2}\, \tilde{\gamma}^{ij}\, \partial_i \kappa\, \partial_j \chi
\\[6pt]
&\quad
- 4 \alpha \alpha_1 \chi\, \tilde{\gamma}^{ij}\,
  \partial_i \kappa\, \partial_j \phi
+ 4 \alpha \alpha_1 \Pi \Xi
\\[6pt]
&\quad
+ \frac{4\alpha}{\chi}\,
  e^{-4\alpha_1 \phi}\,
  \tilde{\gamma}_{ij} B^i E^j .
\end{aligned}
\end{equation}
The equations for the electromgnetic fields, including hyperbolic divergence cleaning become
\begin{equation}
\begin{aligned}
\partial_0 \Psi
&= - \alpha\, \partial_i E^i
\\[3pt]
&\quad
+ \frac{3\alpha}{2\chi}\, E^i \partial_i \chi
+ 2 \alpha_0 \alpha\, E^i \partial_i \phi
\\[3pt]
&\quad
+ \alpha e^{2\alpha_0 \phi}\, B^i \partial_i \kappa
- \alpha \eta_1 \Psi .
\end{aligned}
\end{equation}
\begin{equation}
\begin{aligned}
\partial_0 \Phi
&= \alpha\, \partial_i B^i
- \frac{3\alpha}{2\chi}\, B^i \partial_i \chi
- \alpha \eta_2 \Phi .
\end{aligned}
\end{equation}
\begin{equation}
\begin{aligned}
\partial_0 E^i
&= \alpha K E^i
\\[4pt]
&\quad
+ 2 \alpha \alpha_0 \chi^{1/2}\,
  \epsilon^{jik}\tilde{\gamma}_{kl} B^l \partial_j \phi
- 2 \alpha \alpha_0 E^i \Pi
\\[6pt]
&\quad
+ \alpha e^{2\alpha_0 \phi} \chi^{1/2}\,
  \epsilon^{ijk}\tilde{\gamma}_{kl} E^l \partial_j \kappa
- \alpha e^{2\alpha_0 \phi} \Xi B^i
\\[6pt]
&\quad
- \alpha \chi\, \tilde{\gamma}^{ij} \partial_j \Psi
- \chi^{1/2}\, \epsilon^{jik}\tilde{\gamma}_{kl} B^l \partial_j \alpha
\\[4pt]
&\quad
- \alpha \chi^{1/2}\, \epsilon^{jik} B^l \partial_j \tilde{\gamma}_{kl}
- \alpha \chi^{1/2}\, \epsilon^{jik}\tilde{\gamma}_{kl} \partial_j B^l
\\[4pt]
&\quad
+ \alpha \chi^{-1/2}\, \epsilon^{jik}\tilde{\gamma}_{kl} B^l \partial_j \chi .
\end{aligned}
\end{equation}

\begin{equation}
\begin{aligned}
\partial_0 B^i
&= \alpha K B^i
+ \alpha \chi\, \tilde{\gamma}^{ij} \partial_j \Phi
\\[6pt]
&\quad
- \chi^{1/2}\, \epsilon^{ijk}\tilde{\gamma}_{kl} E^l \partial_j \alpha
+ \alpha \chi^{-1/2}\, \epsilon^{ijk}\tilde{\gamma}_{kl} E^l \partial_j \chi
\\[4pt]
&\quad
- \alpha \chi^{1/2}\, \epsilon^{ijk} E^l \partial_j \tilde{\gamma}_{kl}
- \alpha \chi^{1/2}\, \epsilon^{ijk}\tilde{\gamma}_{kl} \partial_j E^l .
\end{aligned}
\end{equation}

\section{Initial Data}
\label{sec:appx:id}

It is necessary to have good initial data for our EMDA BH binaries.  Of course, exact solutions are not known.  But we can generalize standard approaches for the construction of initial data in vacuum GR to these additional fields on making certain approximations.  We follow closely the approach based on the conformal transverse-traceless decomposition  used with punctures.  Indeed, we extend this approach as it has been incorporated into the {\tt TwoChargedPunctures} code of Bozzola and Paschalidis~\cite{Bozzola2019,ferreira2026electromagneticdualitydegeneracydynamical}.  We mostly follow their notation below.

As usual, we decompose the metric, but using notation slightly modified from the previous appendix.  In particular we define 
%systemsAs a first step, we assume conformally flat space; that is, we decompose the metric as
\begin{equation}
    \gamma_{ij}= \psi^4 \bar{\gamma}_{ij},
\end{equation}
with $\psi$ the conformal factor and the conformal metric assumed to be the flat metric, ${\bar\gamma}_{ij} = \eta_{ij}$.  The latter assumption simplifies the following computations and is a reasonable choice for the systems we study. It is a limiting assumption preventing the construction of highly spinning black holes~\cite{ValienteKroon2004,Price2000}. 
Notwithstanding, Bowen--York black holes can reach spin values of order $\chi \sim 0.9$ \cite{Dain2002}. As our simulations remain below this limit, the approximation should be reasonable.  
While we anticipate the appearance of ``junk radiation'' partly in consequence of this assumption~\cite{Bozzola2021}, we find that it does decay  
as the fields relax to quasi-equilibrium which results in decreased constraint violation over the course of the evolutions. We adopt the maximal slicing condition, $K=0$, and define a conformally rescaled tracefree extrinsic curvature, ${\bar A}_{ij}=\psi^{2}(K_{ij} - \tfrac{1}{3} \gamma_{ij} K)$.  Ignoring the latter's radiative degrees of freedom, we can express it in terms of a vector, $V^i$, as
\begin{equation}
    \bar{A}_{ij} = \partial_i V_j + \partial_j V_i - \frac{2}{3}\eta_{ij}\partial_k V^k.
\end{equation} 

The initial data problem then reduces to determining $V^i$ and $\psi$ for the gravitational pieces.  In particular, the Hamiltonian and momentum constraint equations take the form
\begin{equation}     
\partial_k\partial^k \psi + \tfrac{1}{8} \psi^{-7} \bar{A}_{ij} \bar{A}^{ij} = - 2\pi \psi^5 \rho , \\
\end{equation}
\label{hamiltonian_constraint} 
\begin{equation}
    \partial_k \partial^k V^i + \tfrac{1}{3}\eta^{ij}\partial_j(\partial_k V^k ) = 8 \pi S^i . 
\label{momemtum_constraint}
\end{equation}

The momentum constraint is linear and independent of $\psi$.  This permits dividing the solution between homogeneous (vacuum gravitational) and inhomogeneous (matter) contributions:
\begin{equation}
    V^i = V^i_{\text{GR}} + V^i_{\text{M}}.
\end{equation}
For $V^i_{\text{GR}}$, we use the standard Bowen--York solutions~\cite{Bowen1980} while for the matter contribution, we must solve Eq.~\ref{momemtum_constraint} with 
\begin{equation}
    S^i = 2 \, \Pi \, D^i \phi 
    + 2 e^{-2 \alpha_0 \phi} \epsilon^{ijk} E_j B_k 
    + \tfrac{1}{2}e^{4 \alpha_1 \phi} \, \Xi \, D^i \kappa.
\end{equation}
If we assume a moment of time symmetry and neglect contributions from the magnetic field this source term vanishes.  On imposing asymptotically flat boundary conditions, the solution to the matter part therefore becomes $V^i_{\text{M}}=0$.  

We must also impose the electromagnetic constraints. At this point we  depart somewhat from~\cite{Bozzola2019} because the EMDA version of these constraints are no longer linear, containing rather a nontrivial current that couples the dilaton and axion to the electromagnetic fields: 
\begin{align}
\nabla_i E^i &= 2\alpha_0 E^i \partial_i \phi + e^{2\alpha_0 \phi} B^i \partial_i \kappa \\
\nabla_i B^i &= 0 
\end{align} 
Were the source in the Gauss constraint zero, we would 
follow \cite{Alcubierre2009,Zilhao2012} and perform a conformal rescaling of the fields as
\begin{equation}
    \bar{E}^i = \psi^6 E^i, 
    \quad
    \bar{B}^i = \psi^6 B^i.  
\end{equation}
This results in the sourceless Maxwell constraints taking the form 
\begin{equation}
    \partial_i \bar{E}^i = 0, 
    \quad
    \partial_i \bar{B}^i = 0,
\end{equation}
independent of $\psi$ and therefore solvable separately from the spacetime variables. Since these equations are linear, solutions may be superposed. The advantages of this are significant so we choose to make the approximation that this source term is, in fact, negligible and we solve the EMDA electromagnetic constraints in the way just described.  Of course, this failure to self-consistently solve these constraints necessarily introduces violations.  Partly for this reason, we have introduced hyperbolic divergence cleaning to drive these violations smaller over the course of the evolution.  We have measured these violations on the initial slice. For the simulations considered in this work we confirm that they are consistently small and driven smaller via the constraint damping. The expected damping behavior is demonstrated in Figure~\ref{fig:constraint_divb}. There is an initial violation of the EM constraints, which damps over time.    

The Hamiltonian constraint above, not being linear, must be treated differently.  We assume that we have two black holes charged under the U(1) gauge field as well as possessing scalar dilaton and axion fields.  Following \cite{Bozzola2019}, we approximate this with a conformal factor of the form 
\begin{equation}
    \psi^2 = (1 + u + \bar{\eta} )^2 - \bar{\xi}^2 
\end{equation}
for which we define 
\begin{align}
   \bar{\eta} &= \frac{m_1}{2R_1} + \frac{m_2}{2R_2}, \\
   \bar{\xi}  &= \frac{q_1}{2R_1} + \frac{q_2}{2R_2}, \\ 
   \bar{\kappa} &= 1+ u + {\bar\eta},
\end{align}
and where $m_i$, $q_i$ and $R_i$ here are the masses, electric charges and the coordinate distances from the origin of the BHs.  This is intended to correspond to two electrically charged BHs with the singular part of the conformal factor separated out with $u$ giving finite corrections, including the necessary corrections for the additional scalar fields. 

On rearranging the Hamiltonian constraint in terms of $u$, we find 
\begin{multline}
    \partial_k\partial^k u = -\frac{1}{\bar{\kappa}} \Big(
    \partial_a \bar{\kappa} \partial^a \bar{\kappa} 
    - \partial_a \bar{\xi} \partial^a \bar{\xi} 
    - \partial_a \psi \partial^a \psi \\
    + \tfrac{1}{8} \psi^{-6} \bar{A}_{ij} \bar{A}^{ij}
    + 2 \pi \psi^{6} \rho
    \Big),
\end{multline}
with
\begin{equation}
\begin{split}
    \rho \equiv\;& \bigl[ \Pi^2 + (D\phi)^2 \bigr]
    + e^{-2 \alpha_0 \phi}\big[ E_a E^a +  B_a  B^a\big] \\
    &+ \tfrac{1}{4} e^{4 \alpha_1 \phi}\big[\Xi^2 + (D\kappa)^2\big].
\end{split}
\end{equation}
Using simplifying assumptions similar to above, namely a moment of time symmetry and neglecting the magnetic field, $\rho$ reduces to
\begin{equation}
    \rho = (D\phi)^2 + e^{-2 \alpha_0 \phi} \, E_a E^a + \tfrac{1}{4} e^{4\alpha_1 \phi} \, (D\kappa)^2 .
\end{equation}
Again, this follows \cite{Bozzola2019} quite closely, except that we have now included dilaton and axion contributions in $\rho$. Specifically, we employ a Reissner--Nordström electric field for the electric component and add a dilaton field for a single, spinless black hole. In practice, the axion is initialized to zero and allowed to relax to a quasi-equilibrium state during an evolution.  

The electric field is initially set to be:
\begin{equation}
     E^i = \psi^{-6} \Bigl[ \frac{q_1}{R_1^2} + \frac{q_2}{R_2^2} \Bigr].
\end{equation}
while the dilaton fields are initially set as:
\begin{align}
\phi &= \phi_1(r_1,\theta_1) + \phi_2(r_2, \theta_2) 
\end{align}
where Eq.~\ref{Kerr-Sen_dilaton} gives the form for a dilaton field for a single Kerr--Sen black hole (albeit in Boyer--Lindquist type coordinates) for our current theory parameters of $(\alpha_0,\alpha_1) = (1,1)$. 

\section{Wave Extraction}  
\label{sec:appx:wave_extraction}

The GW signals are given by the Newman--Penrose scalar derived from the Weyl tensor as:
\begin{equation}
    \Psi_4 = C_{abcd} k^{a} \bar{m}^{b} k^{c} \bar{m}^{d}.
\end{equation}
where $k^a$ and $\bar{m}^b$ are members of a null tetrad defined as 
\begin{equation}
\begin{gathered}
l^{a} = \frac{1}{\sqrt{2}}\left(n^{a} + u^{a}\right),\quad
k^{a} = \frac{1}{\sqrt{2}}\left(n^{a} - u^{a}\right),\\
m^{a} = \frac{1}{\sqrt{2}}\left(v^{a} + i w^{a}\right).
\end{gathered}
\end{equation}
This null tetrad is constructed in terms of orthonormal spatial vectors $u^a, v^b, w^c$ which asymptotically behave like the standard radial, polar and azimuthal unit vectors, respectively, and which are normal to the timelike unit vector $n^{a}$.  
 
The extraction of the outgoing electromagnetic radiation is given by two scalar quantities calculated in the Newman--Penrose formalism \cite{Newman1962,Zilhao2014}.  
These two radiative quantities are defined as:
\begin{align}
\Phi_1 &= \frac{1}{2} F_{a b}
\left(l^{a} k^{b} + \bar{m}^{a} m^{b}\right),\\
\Phi_2 &= F_{ab} \bar{m}^{a} k^{b}.
\end{align}
The outgoing radiative quantities for the dilaton and axion are defined in a similar way, namely 
\begin{equation}
    \bm{\varphi} = l^a \nabla_a \phi, \quad
    \bm{\vartheta} = l^a \nabla_a \kappa.
\end{equation}

At a given extraction radius $R_{\mathrm{ex}}$, we perform a multipolar decomposition by projecting each field onto spin-weighted spherical harmonics with the appropriate spin weight (taking values $s=-2,-1$ or $0$ depending on the field):
\begin{align}
\Psi_4(t,\theta,\phi) &= \sum_{l,m} \psi^{lm}(t)\, Y^{-2}_{lm}(\theta,\phi), \label{eq:psi4_decomp} \\
\Phi_1(t,\theta,\phi) &= \sum_{l,m} \Phi^{lm}_{1}(t)\, Y^{0}_{lm}(\theta,\phi), \label{eq:xi1_decomp} \\
\Phi_2(t,\theta,\phi) &= \sum_{l,m} \Phi^{lm}_{2}(t)\, Y^{-1}_{lm}(\theta,\phi), \label{eq:xi2_decomp} \\
{\bm{\varphi}}(t,\theta,\phi) &= \sum_{l,m} {\bm{\varphi}}^{lm}(t)\, Y^{0}_{lm}(\theta,\phi), \label{eq:phi_decomp} \\
\bm{\vartheta}(t,\theta,\phi) &= \sum_{l,m} \bm{\vartheta}^{lm}(t)\, Y^{0}_{lm}(\theta,\phi). \label{eq:kappa_decomp}
\end{align}

We can calculate the radiated energy from the gravitational, electromagnetic, axion, and dilaton fields as follows \cite{Zilhao2012}:
\begin{align}
F_{\mathrm{GW}}
&= \frac{dE_{\mathrm{GW}}}{dt}
= \lim_{r \to \infty} \frac{r^2}{16\pi}
\sum_{l,m} \left| \int_{-\infty}^{t} dt'\, \psi^{lm}(t') \right|^2, \\
F_{\mathrm{EM}}
&= \frac{dE_{\mathrm{EM}}}{dt}
= \lim_{r \to \infty} \frac{r^2}{4\pi}
\sum_{l,m} \left| \Phi^{lm}_2(t) \right|^2 , \\ 
F_{\mathrm{D}}
&= \frac{dE_{\mathrm{D}}}{dt}
= \lim_{r \to \infty} \frac{r^2}{4\pi}
\sum_{l,m} \left| \bm{\varphi}^{lm}(t) \right|^2 , \\
F_{\mathrm{A}}
&= \frac{dE_{\mathrm{A}}}{dt}
= \lim_{r \to \infty} \frac{r^2}{4\pi}
\sum_{l,m} \left| \bm{\vartheta}^{lm}(t) \right|^2 . 
\end{align}

\section{Performance of the Code}
\label{sec:code_performance}
\subsection{Constraint quantities}

\begin{figure}[tbp]
  \centering
  \includegraphics[width=0.92\columnwidth]{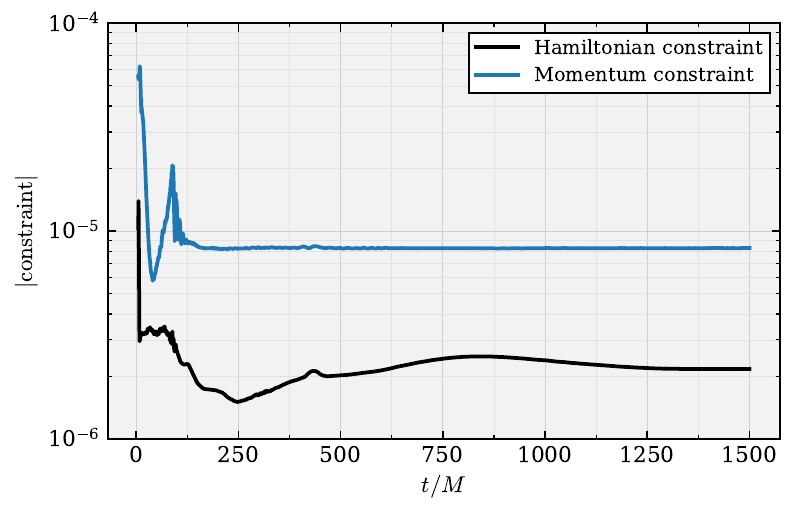}
  \caption{
  The Hamiltonian constraint $\mathcal{H}$ and summed momentum constraint
  violation measured over the full simulation. The $L_2$ norm is taken with respect to the number of points  The constraints remain
  controlled throughout the evolution, including the merger phase.
  }
  \label{fig:constraint_ham_mom}
\end{figure}

\begin{figure}[tbp]
  \centering
  \includegraphics[width=0.92\columnwidth]{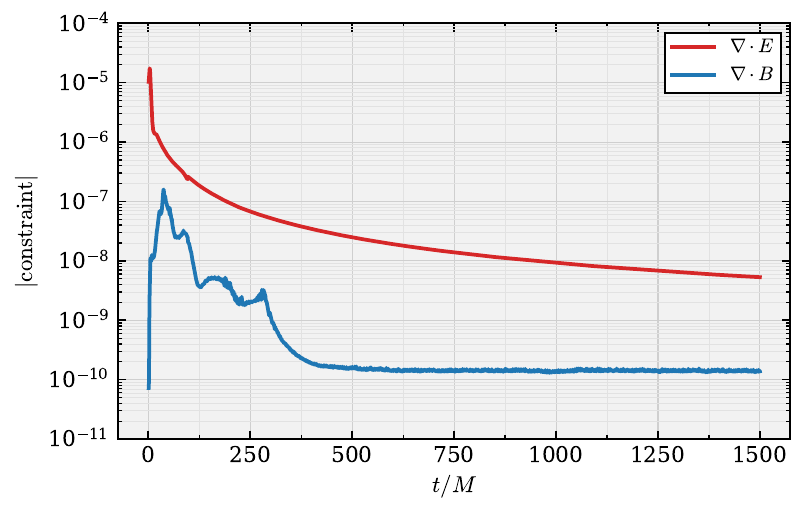}
  \caption{
  The volume weighted $L_2$ norms of the electromagnetic constraints, namely the no-monopole constraint that involves the divergence of the magnetic field, $\nabla\cdot B$, and the Gauss constraint that invokes the divergence of the electric field.  Both are measured over the full simulation. 
  }
  \label{fig:constraint_divb}
\end{figure}

We monitor the constraint violations throughout the evolution.
In particular, we compute the $L_2$ norms of the Hamiltonian and momentum
constraints, together with the electromagnetic constraints.
When evaluating these norms, we excise small neighborhoods around the punctures
to avoid unduly weighting them by these regions. Representative
constraint violations for the full simulation are shown in
Figs.~\ref{fig:constraint_ham_mom} and \ref{fig:constraint_divb}.

% ---------------------------------------------------------
% Constraint violations over the full simulation
% ---------------------------------------------------------

A full convergence study for binary black hole evolutions is computationally expensive and is therefore not attempted here. Instead, we assess the behavior of the code by varying two parameters that directly affect the adaptive mesh refinement: the wavelet tolerance, $\varepsilon$, and the maximum refinement level, $\ell_{\rm max}$. The wavelet tolerance is the threshold value in our WAMR scheme on the  coefficients of the wavelet expansion that determines whether refinement or coarsening occurs.  The maximum refinement level sets the largest number of refinement levels that can be obtained in a simulation.  We confirm that for both, making these refinement criteria more stringent leads to a corresponding reduction in the constraint violations.

The tests done here use the first \(20M\) of a head-on evolution of two equal-mass, charged black holes. The black holes have an initial coordinate separation of $16M$, individual masses of \(m_1=m_2=0.50\), and charge-to-mass ratios \(q_1/m_1=q_2/m_2=0.1\). Both black holes have dimensionless spin \(\chi=0.4\) and are initially at rest. In computing the constraint diagnostics, we again exclude a small neighborhood of each puncture from calculation of the norms.
Throughout this section, the constraint measure shown is the sum of the volume weighted Hamiltonian constraint and the three volume weighted momentum constraints. This provides a single diagnostic for the overall constraint violation during the evolution.

\begin{figure}[tbp]
  \centering
  \includegraphics[width=0.92\columnwidth]{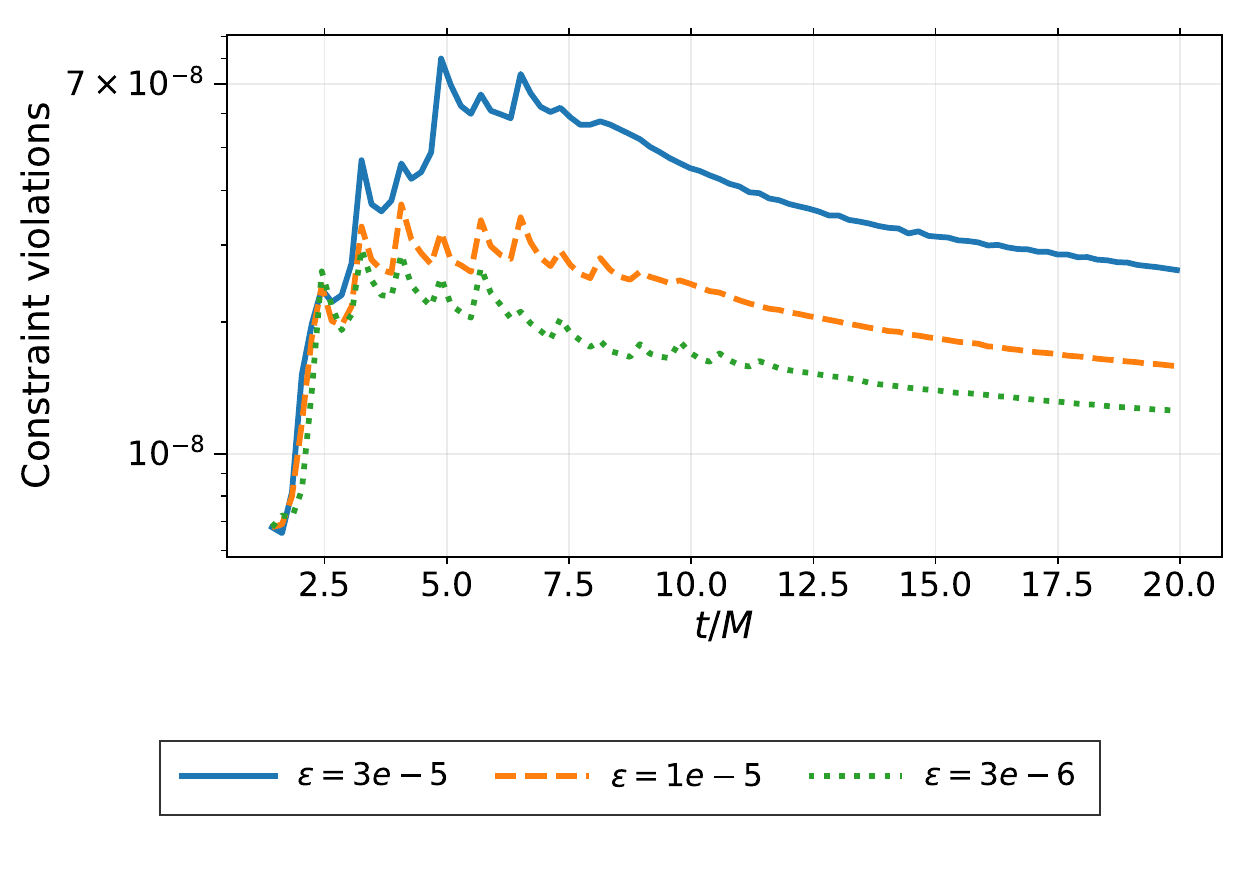}
  \caption{
  The sum of the volume weighted Hamiltonian and momentum constraint violations
  over the first \(20M\) of a head-on evolution of two equal-mass charged
  black holes with charge-to-mass ratio \(q/m=0.1\) and dimensionless spin
  \(\chi=0.4\). The simulations use wavelet tolerances of $\varepsilon \in \{3\times10^{-5}, 1\times10^{-5}, 3\times10^{-6} \}$. These simulations were done with a maximum refinement level of $\ell_{\rm max} = 15$.
  Lowering the wavelet tolerance improves the resolution of dynamically
  important regions and reduces the overall constraint violation.
  }
  \label{fig:constraint_wavelet_tolerance}
\end{figure}

Comparisons of simulations with respect to different wavelet tolerances are shown in Figure~\ref{fig:constraint_wavelet_tolerance}.  Decreasing $\varepsilon$ allows the grid to refine more aggressively in regions with larger truncation error.  For the three different wavelet tolerances of $\varepsilon \in \{3\times10^{-5}, 1\times10^{-5}, 3\times10^{-6}\}$ we can see that the constraints are indeed reduced as $\varepsilon$ decreases.  

\begin{figure}[tbp]
  \centering
  \includegraphics[width=0.92\columnwidth]{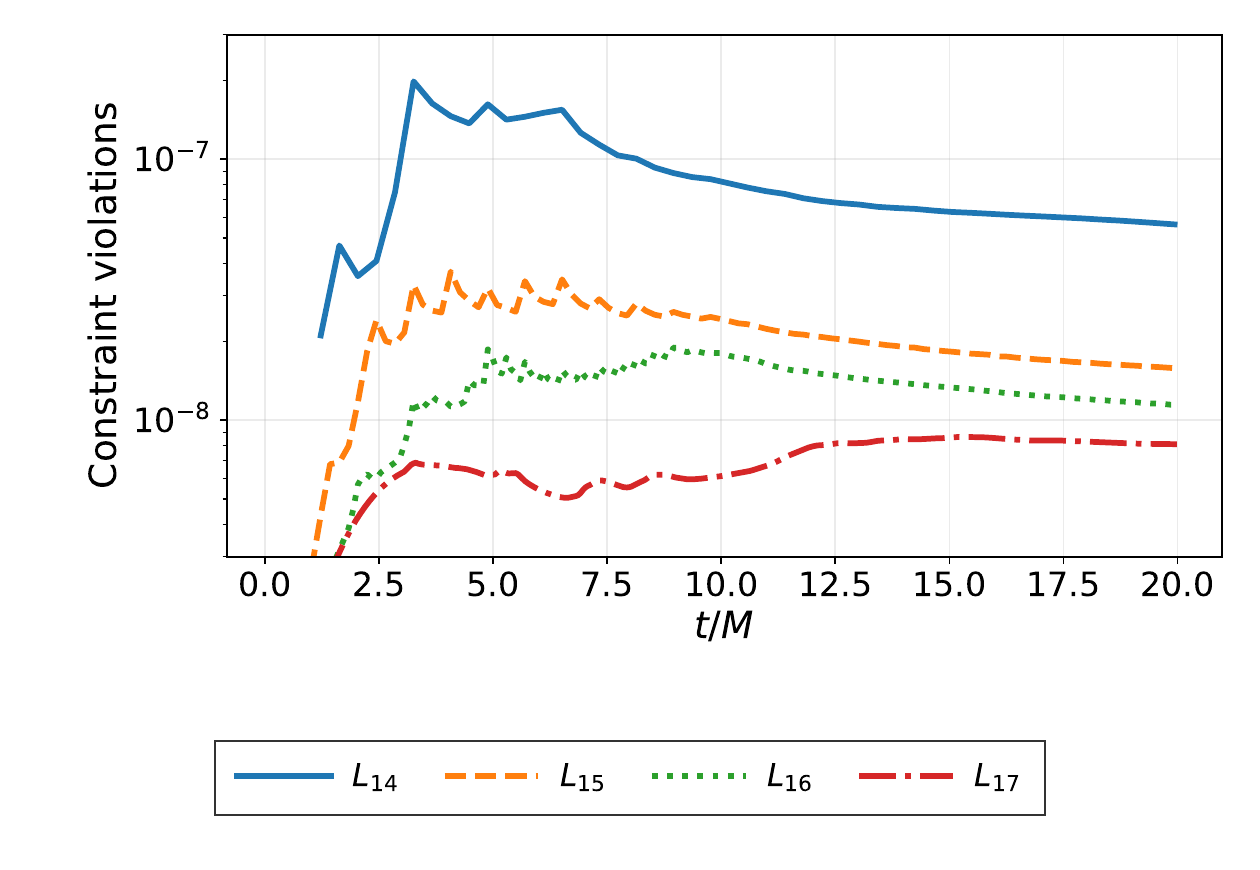}
  \caption{
  The sum of the volume weighted Hamiltonian and momentum constraint violations over the first \(20M\) of a head-on evolution of two equal-mass charged black holes with charge-to-mass ratio \(q/m=0.1\) and dimensionless spin  \(\chi=0.4\) each. The simulations compare maximum refinement levels $\ell_{\rm max} \in \{14, 15, 16, 17\}$ corresponding to finest grid spacings of $h \in \{ 0.0325,0.0162,0.00813,0.00407 \}$, respectively. The wavelet tolerance is set here to $\varepsilon = 1\times 10^{-5}$.  Increasing the maximum refinement level reduces the overall constraint violation by allowing additional grid refinement in under-resolved regions.
  }
  \label{fig:constraint_maxdepth}
\end{figure}

Comparing simulations according to different maximum refinement levels is another means of changing the resolution in an AMR setting.  Figure~\ref{fig:constraint_maxdepth} compares simulations with maximum refinement levels of $\ell_{\rm max} \in \{14, 15, 16, 17\}$ corresponding to finest grid spacings of $h \in \{ 0.0325, 0.0162, 0.00813, 0.00407 \}$, respectively. Increasing the allowed maximum depth provides additional resolution near the black holes and in regions where the fields develop sharp spatial structures. The resulting decrease in the constraints provides further evidence that the simulations improve with increased resolution.

\section{Summary of Tests Done}
\label{sec:appx:sim_tab}
In addition to some of the tests mentioned in the main body of this paper, we have also performed a broader set of simulations to test how the late-time dilaton and axion fields depend on the initial charges and spins of the binary components. In particular, we considered Kerr--Sen binaries with same-sign and opposite-sign charges, spin-aligned and spin-anti-aligned configurations, as well as mergers involving neutral GR black holes. For binaries with opposite charges, the net charge vanishes and the remnant settles to a Kerr black hole, with neither a persistent dilaton nor axion field. For binaries with same-sign charges but oppositely aligned spins, the axion field decays, while the dilaton field remains nontrivial, corresponding to an EMD-like remnant. We further find that when charged, spinless black holes are merged with Kerr black holes, both axion and dilaton fields persist. Finally, we find that mergers of initially nonscalarized Kerr--Newman black holes produce remnants with both dilaton and axion hair. The configurations considered are summarized in Table~\ref{tab:bh_hair_combined}, where a green check mark indicates that the corresponding field remains persistent and nontrivial at late times.

\FloatBarrier
\twocolumngrid  % ensure we are in 2-column before switching
\onecolumngrid

\FloatBarrier
% ----- single-column merged table -----
\begin{center}
{\label{tab:bh_hair_combined}%
Binary black hole configurations }
\begingroup
\footnotesize
\setlength{\tabcolsep}{8pt}
\renewcommand{\arraystretch}{1.1}

\begin{ruledtabular}

% ---------- FIRST SECTION ----------
\textbf{(a) Kerr--Sen Same-Sign Charge and Spin-Aligned Black Hole Tests} \\[0.4em]
\begin{tabular}{lllcc}
$(\chi_1, q_1)$ & $(\chi_2, q_2)$ & \textrm{Dilaton $\boldsymbol{\phi}$} & \textrm{Axion $\boldsymbol{\kappa}$} \\ \hline
$\chi_1 = 0.10$, $q_1 \in \{0.0624, 0.125, 0.25, 0.50\}$ &
$\chi_2 = 0.10$, $q_2 \in \{0.0624, 0.125, 0.25, 0.50\}$ &
\cmark & \cmark \\[0.3em]
$\chi_1 = 0.20$, $q_1 \in \{0.0624, 0.125, 0.25, 0.50\}$ &
$\chi_2 \in \{0.10, 0.20\}$, $q_2$ as above &
\cmark & \cmark \\[0.3em]
$\chi_1 = 0.40$, $q_1 \in \{0.0624, 0.125, 0.25, 0.50\}$ &
$\chi_2 \in \{0.10, 0.20, 0.40\}$, $q_2 \in \{0.0624, 0.125, 0.25, 0.50\}$ &
\cmark & \cmark \\
\end{tabular}

\vspace{1.2em} % spacing between subtables

% ---------- SECOND SECTION ----------
\textbf{(b) Collisions with Other Black Holes} \\[0.4em]
\begin{tabular}{llcc}
\textrm{Black Hole 1 $(q,\chi)$} & \textrm{Black Hole 2 $(q,\chi)$} & $\boldsymbol{\phi}$ & $\boldsymbol{\kappa}$ \\
\hline
Kerr--Sen $(0.5,\,0.4)$   & Kerr--Sen $(0.5,\,0.4)$   & \cmark & \cmark \\
Kerr--Sen $(0.5,\,0.4)$   & Kerr $(0.0,\,0.4)$        & \cmark & \cmark \\
Kerr--Sen $(0.5,\,0.4)$   & Kerr--Sen $(-0.5,\,0.4)$  & \xmark & \xmark \\
Kerr--Sen $(0.5,\,-0.4)$  & Kerr--Sen $(0.5,\,0.4)$   & \cmark & \xmark \\
EMD $(0.5,\,0.0)$          & Kerr $(0.0,\,0.4)$        & \cmark & \cmark \\
KN $(0.5,\,0.4)^{\dagger}$& KN $(0.5,\,0.4)^{\dagger}$& \cmark & \cmark \\
\end{tabular}

\end{ruledtabular}
\endgroup

\vspace{0.5em}
\footnotesize
\noindent
$^{\dagger}$ These configurations were initialized as nonscalarized
Kerr--Newman black holes.

Abbreviations: EMD denotes Einstein--Maxwell--dilaton, corresponding here to
a charged black hole with dilaton hair; KN denotes Kerr--Newman, corresponding
to a black hole with charge and spin; and Kerr--Sen denotes a rotating charged
black hole with dilaton and axion hair.
\end{center}

% ----- end single-column content -----

% return to two-column mode
\twocolumngrid

\FloatBarrier

\bibliography{references}  % Produces the bibliography via BibTeX

\end{document}